\documentclass[%
 reprint,
superscriptaddress,
showpacs,preprintnumbers,
 amsmath,amssymb,
 aps,
pra,
]{revtex4-1}

\usepackage{graphicx}
\usepackage{dcolumn}
\usepackage{bm}
\usepackage{xcolor}
\usepackage{soul}

\begin{document}

\preprint{APS}

\title{Highly entangled-photon pairs generated from the biexciton cascade transition in a quantum dot-metal nanoparticle hybrid system}%

\author{T. Moradi}%
 \affiliation{Department of Physics, Faculty of Science, University of Isfahan, Hezar Jerib Str., Isfahan 81746-73441, Iran.}
\author{M. Bagheri Harouni}%
 \email{m-bagheri@phys.ui.ac.ir}
\affiliation{Department of Physics, Faculty of Science, University of Isfahan, Hezar Jerib Str., Isfahan 81746-73441, Iran.}%
\affiliation{Quantum Optics Group, Department of Physics, Faculty of Science, University of Isfahan, Hezar Jerib Str., Isfahan 81746-73441, Iran}%
\author{M. H. Naderi}%
\affiliation{Department of Physics, Faculty of Science, University of Isfahan, Hezar Jerib Str., Isfahan 81746-73441, Iran.}%
\affiliation{Quantum Optics Group, Department of Physics, Faculty of Science, University of Isfahan, Hezar Jerib Str., Isfahan 81746-73441, Iran}%

\date{\today}

\begin{abstract}
The entanglement between photon pairs generated from the biexciton
cascade transition in a semiconductor quantum dot located in the
vicinity of a metal nanoparticle is theoretically investigated. In
the model scheme, the biexciton-exciton and exciton-ground state
transitions are assumed to be coupled to two principal plasmon modes
of orthogonal polarizations. For a broad spectral window, because
the horizontal and vertical spectra are overlapped, the biexciton
and exciton photons are degenerate in energy. This allows us to
overcome the natural splitting between the intermediate exciton
states. Moreover\texttt{\emph{,}} the degree of entanglement depends
on the geometrical parameters of the system. i.e.\texttt{\emph{,}}
the radius of the metal nanoparticle and the distance between the
quantum dot and the nanoparticle. The results reveal that such a
hybrid system profoundly modifies the photon entanglement even in
the absence of strong coupling between the emitter and the metal
nanosphere.
\end{abstract}

\maketitle


\section{Introduction}
    \par
\indent Quantum entanglement, as the most remarkable characteristic
of composite quantum systems, is the essence of quantum theory as
suggested by Schr$\ddot{o}$dinger \cite{1} and as emphasized by the
counterintuitive and distinctive aspects of quantum mechanics
against the classical world. It is at the heart of the
Einstein-Podolsky-Rosen paradox \cite{2}, of Bell's inequalities
\cite{3}, and of the so-called "quantum nonlocality" \cite{4}. Apart
from its central role in testing the fundamental aspects of quantum
mechanics, entanglement has proven to be an essential ingredient of
many protocols for quantum information processing such as quantum
teleportation \cite{5,6}, quantum cryptography \cite{7}, and quantum
computation \cite{8}. In addition, quantum entanglement can be
employed for high-precision spectroscopy \cite{9} and quantum
simulation \cite{10}. Therefore, during the last two decades or so,
considerable interest has been focused on the controlled, generation
and preservation of entanglement.\par To date, various theoretical
schemes and experimental demonstrations have been carried out on the
generation of entangled states of massive particles, such as atoms
\cite{11}, trapped ions \cite{12}, and atomic ensembles \cite{13}.
However, the entangled states of such massive particles are
extremely fragile and sensitive to any kind of fluctuations or other
decoherence processes. On the other hand, the entanglement of
photons has been proven to be easier to create and maintain than
that of material particles \cite{14}. There currently exist a
variety of schemes for generation of entangled photons, e.g., based
on atomic cascade decay \cite{15,17}, parametric down-conversion in
optical nonlinear crystals \cite{18}, spontaneous four-wave mixing
in an optical fiber \cite{19}, and radiative biexciton cascade in
semiconductor quantum dots (QDs) \cite{20,20-1,20-2,20-3,20-4,20-5,21,22,23}.\par The radiative
biexciton cascade transition in a single semiconductor QD provides a
source of entangled photons \cite{25,26,27,28,29-1,30,30-1,20-4}. A biexciton cascade can be described as a four-level system composed
of a biexciton state ($\left| u \right\rangle $)\texttt{\emph{,}}
two bright intermediate exciton levels ($\left| x \right\rangle ,$
$\left| y \right\rangle $)\texttt{\emph{,}} and a ground state
($\left| g \right\rangle $)  \cite{20}. The biexciton state can
spontaneously decay to the ground state through two intermediate
exciton states by the emission of a pair of photons; biexciton
photon and exciton photon resulting from transitions $ \left| u
\right\rangle \to \left| x \right\rangle (\left| y \right\rangle ) $
and $\left| x \right\rangle (\left| y \right\rangle )\to \left| g
\right\rangle $, respectively. The polarization of emitted photons
is determined by the spin of the intermediate exciton states. In the
case of an idealized QD, the intermediate bright exciton states are
degenerate in energy, resulting in two indistinguishable decay
paths. Thus the biexciton radiative decay can lead to the generation
of polarization-entangled photon pairs. However, in practice, the
asymmetry of the QD shape leads to a fine structure splitting (FSS)
between the intermediate exciton states \cite{31,32}. The
nonvanishing FSS encodes the "which-path" information and, as the
result, the photon polarizations are only classically correlated
rather than entangled. The radiation of entangled photon pairs from
a QD therefore relies on the reduction of the FSS to zero. To
achieve the photon pairs with the significant degree of entanglement
the FSS between the intermediate exciton states should be smaller
than the exciton radiative linewidth of typically $1\mu $eV
\cite{22}. Several methods have been employed to tune this FSS,
including the use of external magnetic fields \cite{29}, external
electric fields \cite{30}, external uniaxial stress \cite{33}, and
   electro-elastic fields \cite{20-4}.
Other approaches include spectral filtering \cite{22},  selection of
QDs with low FSSs \cite{35}, and using a single QD strongly coupled
to a planar photonic crystal \cite{24,27,28}. \par In the present
paper, we theoretically investigate the polarization entanglement of
the photon pairs emitted from a single semiconductor QD with cascade
configuration in the vicinity of a metal nanoparticle (MNP). Our
main purpose is twofold. On the one hand, we aim at proposing a
novel approach to overcome FSS for generating highly entangled
photon pairs in the QD-MNP hybrid system, and on the other hand, we
intend to explore whether the degree of entanglement can be
controlled by the geometrical parameters of the hybrid system. \par
The interaction of metal nanostructures with light can give rise to
collective excitations, known as localized surface plasmon (LSP)
resonances \cite{36}. The large mismatch between the wavelength of
light and the size of single emitters ensures that the light-matter
interaction is inherently weak \cite{37,38}. Several methods have
been proposed to strengthen the interaction between light and matter
based on decreasing the effective mode volume and increasing the
Rabi frequency. It has been shown \cite{39} that the strong coupling
between emitters close to MNPs is possible due to the small mode
volume of highly confined evanescent field associated with surface
plasmons. By confining light using localized surface plasmone, the
local density of states alters significantly \cite{40,41}. As a
consequence, the light-matter interaction can be significantly
enhanced. Although the confined plasmonic modes couple very strongly
with matter for large ohmic losses, unfortunately, it is not easy to
enter the strong-coupling regime in plasmonic systems \cite{42}.
Various quantum optical features of MNP coupling to a variety of QD
configurations have been extensively investigated within different
approaches for both the weak and strong coupling regimes
\cite{40,41,42,43,44,45,46,47,48}\par
In this paper, we study the
manipulation of entanglement of the photon pairs emitted from the
biexciton cascade coupled to the surface plasmon modes via the
changing geometry and other relevant physical parameters of the
hybrid system under consideration. In general, there are two
fundamental sources of noise\texttt{\emph{,}} i.e.\texttt{\emph{,}}
thermal and quantum noises. Thermal noise is usually omnipresent and
quantum noise is unavoidable. It is apparent that modes having
frequencies much higher than ${{k}_{B}}T$ are rarely excited
thermally. In a  hybrid system composed of the MNP and QD, the
typical values of energies are higher than ${{k}_{B}}T$. Therefore,
we assume that the system is treated as one would treat it at zero
temperature \cite{41}. Since a MNP acts as a dispersive and
absorbing medium, our treatment is based on the macroscopic
quantization of the electromagnetic fields which utilizes the
classical dyadic Green's function and includes quantum noise sources
\cite{49,50}. The paper is structured as follows. In Sec.II, we
first introduce the theoretical model of a single QD coupled to a
metal nanosphere with two orthogonal principal plasmon modes in the
weak-coupling regime. Then, we determine the probability amplitudes
for the emission of two orthogonal-polarization photons emitted from
the QD in the long time limit. In Sec. III we consider the spectral
functions as well as the polarization entanglement of the generated
photon pairs. Section IV is devoted to the discussion of the
numerical results. In particular, we explore the effects of the
radius of the MNP and the QD-MNP separation distance on the degree
of entanglement. Finally, our conclusions are summarized in Sec. V.

\section{DESCRIPTION OF THE HYBRID SYSTEM }
As is shown in Fig. (1)\texttt{\emph{,}} the physical system we
consider consists of a QD located at distance h from the surface of
a metal nanosphere of radius R and frequency- dependent permittivity
$\varepsilon (\omega )$.

    \begin{figure}
        \includegraphics[width=\linewidth]{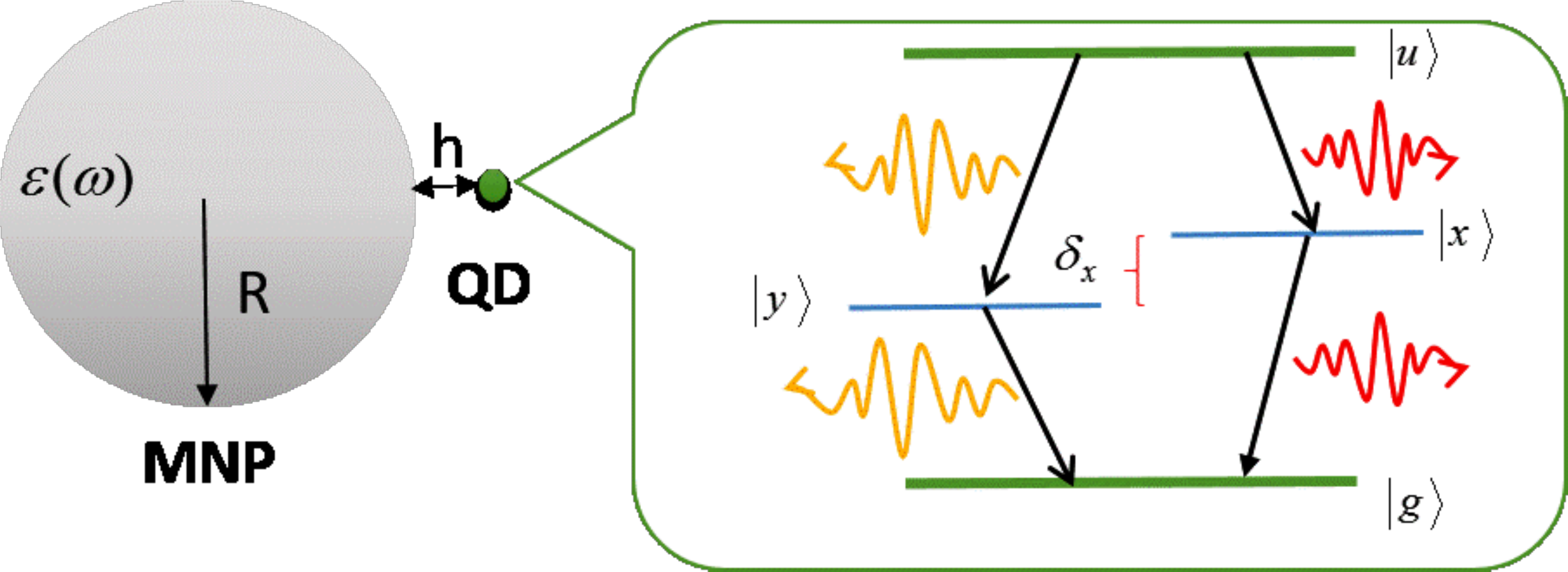}
        \caption{Schematic of the QD-MNP hybrid system studied in the paper.
        A QD located at distance h from the surface of a MNP. The discrete
        energy states of the QD are the biexciton state $\left| u \right\rangle $,
        the exciton states $\left| x \right\rangle {{,}_{{}}}\left| y \right\rangle $,
        and ground state$\left| g \right\rangle $. The two bare exciton levels are
        splitted by an energy ${{E}_{x}}-{{E}_{y}}={{\delta }_{x}}$.}
        \label{fig:1}
    \end{figure}
\par The energies of the biexciton and intermediate exciton states are
${{E}_{u}}=2{{E}_{0}}-{{\Delta }_{xx}}$, ${{E}_{x}}$ and
${{E}_{y}}$, respectively. Here ${{\Delta }_{xx}}$ is the binding
energy of the biexciton and denotes the biexciton energy shift due
to the exciton-exciton interaction. Moreover\texttt{\emph{,}}
${E_{0}}$ is the mean value of the energies of the intermediate
states. It should be note that the ground state energy is selected
equal to zero. The cascade emission process starts from the
biexciton state passing through the intermediate exciton states,
$\left| x \right\rangle $ or $\left| y \right\rangle $, to reach the
ground state. The two bare exciton levels have a FSS
${{E}_{x}}-{{E}_{y}}={{\delta }_{x}}$. We assume that the both
biexciton-exciton and exciton-ground state transitions are coupled
to two principal plasmon modes of orthogonal polarizations. In the
model under study, we consider a general realistic scenario where
each subsystem is coupled to its own reservoir so that the total
system has three separated reservoirs\texttt{\emph{,}}
i.e.\texttt{\emph{,}} two plasmonic reservoirs and a QD reservoir.
\par In this study, we are dealing with the electrical pump which is usually used in the direct gap semiconductors. The self-assembled QD, which is embedded in a wetting layer (WL), is photo-excited and, consequently, electrons in the WL conduction band and holes in the WL valence band are created. These carriers subsequently relax into the QD and occupy the discrete energy following Pauli’s Principle. Although relaxed, they still can interact with the WL via multi-photon processes.\cite{59,60,61}
Using this mechanism, we investigate the biexciton cascade and formation of entangled photons from a single QD.
\subsection{System Hamiltonian   }

The total Hamiltonian of the system can be written as:
    \begin{align}\label{eq:1}
    & {\bm{\hat{H}}_{I}}(t)=\\
    &\bm{\hat{H}}_{Plas-QD}^{I}(t)+
    \bm{\hat{H}}_{Plas-\operatorname{Re}s}^{I}(t)+\bm{\hat{H}}_{QD-\operatorname{Re}s}^{I}(t)\texttt{
        \emph{,} }\nonumber
    \end{align}
where $\bm{\hat{H}}_{Plas-QD}^{I}(t)$ describes the interaction
between the plasmonic modes and QD,
$\bm{\hat{H}}_{Plas-\operatorname{Re}s}^{I}(t)$ denotes the coupling
of plasmonic modes with their reservoirs, and
$\bm{\hat{H}}_{QD-\operatorname{Re}s}^{I}(t)$ refers to the
interaction between the QD and its reservoir. In the interaction
picture and under the rotating-wave approximation the explicit form
of these Hamiltonians can be written as
    \begin{subequations}
        \begin{flalign}\label{eq:2}
        & \bm{\hat{H}}_{Plas-QD}^{I}(t)=\hbar [g_{ex}^{x}|x\rangle\langle g|\bm{\hat{a}}_{p}^{x}{{e}^{i\Delta _{p}^{x}t}}\nonumber\\
        & +g_{bx}^{x}|u\rangle\langle x|\bm{\bm{\hat{a}}}_{p}^{x}{{e}^{i({{\omega }_{ux}}-\omega _{p}^{x})t}}
        +g_{ex}^{y}|y\rangle\langle g|\bm{\hat{a}}_{p}^{y}{{e}^{i\Delta _{p}^{y}t}} \nonumber\\
        & +g_{bx}^{y}|u\rangle\langle y|\bm{\hat{a}}_{p}^{x}{{e}^{i({{\omega }_{uy}}-\omega _{p}^{y})t}}]+H.C.\texttt{\emph{,} } \\
        & \bm{\hat{H}}_{Plas-\operatorname{Re}s}^{I}(t)=\hbar [\int\limits_{0}^{\infty }{d{{\omega }_{m}}({{\Omega }_{xm}}\bm{\hat{a}}{{_{p}^{x}}^{\dagger }}\bm{\hat{f}}({{\omega }_{m}})} \nonumber\\
        & {{e}^{i(\omega _{p}^{x}-{{\omega }_{m}})t}})+\int\limits_{0}^{\infty }{d{{\omega }_{m}}({{\Omega }_{ym}}\bm{\hat{a}}{{_{p}^{y}}^{\dagger }}\bm{\hat{f}}({{\omega }_{m}})} \nonumber\\
        & {{e}^{i(\omega _{p}^{y}-{{\omega }_{m}})t}})]+H.C.\texttt{\emph{,} }  \\
        & \bm{\hat{H}}_{QD-\operatorname{Re}s}^{I}(t)=-|x\rangle\langle g|\int_{0}^{\infty }{d{{\omega }_{q}}(}{{d}_{gx}} \nonumber\\
        & i\sqrt{\frac{\hbar }{\pi {{\varepsilon }_{0}}}}\int{{{d}^{3}}\bm{{r}'}\frac{{{\omega }_{q}}^{2}}{{{c}^{2}}}}\sqrt{{{\varepsilon }_{I}}(\bm{{r}'},{{\omega }_{q}})} \nonumber\\
        & \bm{G}({{{\bm{r}}}_{d}},\bm{{r}'},{{\omega }_{q}})\bm{\hat{f}}(\bm{{r}'},{{\omega }_{q}}){{e}^{i({{\omega }_{x}}-{{\omega }_{q}})t}})- \nonumber\\
        & |u\rangle\langle x|\int_{0}^{\infty }{d{{\omega }_{k}}}{{d}_{ux}}i\sqrt{\frac{\hbar }{\pi {{\varepsilon }_{0}}}}\int{{{d}^{3}}\bm{{r}''}}\nonumber \\
        & (\frac{{{\omega }_{k}}^{2}}{{{c}^{2}}}\sqrt{{{\varepsilon }_{I}}(\bm{{r}''},{{\omega }_{k}})}\bm{G}({{{\bm{r}}}_{d}},\bm{{r}''},{{\omega }_{k}})\bm{\hat{f}}(\bm{{r}''},{{\omega }_{k}}) \nonumber\\
        & {{e}^{i({{\omega }_{ux}}-{{\omega }_{k}})t}})-|y\rangle\langle g|\int_{0}^{\infty }{d{{\omega }_{q}}}{{d}_{gy}}i\sqrt{\frac{\hbar }{\pi {{\varepsilon }_{0}}}} \nonumber\\
        & \int{{{d}^{3}}\bm{{r}'}(\frac{{{\omega }_{q}}^{2}}{{{c}^{2}}}}\sqrt{{{\varepsilon }_{I}}(\bm{{r}'},{{\omega }_{q}})}\bm{G}({{{\bm{r}}}_{d}},\bm{{r}'},{{\omega }_{q}}) \nonumber\\
        & \bm{\hat{f}}(\bm{{r}'},{{\omega }_{q}}){{e}^{i({{\omega }_{y}}-{{\omega }_{q}})t}})-|u\rangle\langle y|\int_{0}^{\infty }{d{{\omega }_{k}}(} \nonumber\\
        & {{d}_{uy}}i\sqrt{\frac{\hbar }{\pi {{\varepsilon }_{0}}}}\int{{{d}^{3}}\bm{{r}''}\frac{{{\omega }_{k}}^{2}}{{{c}^{2}}}}\sqrt{{{\varepsilon }_{I}}(\bm{{r}''},{{\omega }_{k}})} \nonumber\\
        & \bm{G}({{{\bm{r}}}_{d}},\bm{{r}''},{{\omega }_{k}})\bm{\hat{f}}(\bm{{r}''},{{\omega }_{k}}){{e}^{i({{\omega }_{uy}}-{{\omega }_{k}})t}})+H.C.\texttt{\emph{,} }
        \end{flalign}
    \end{subequations}
where,  ${{\omega }_{ux(y)}}=({{\omega }_{u}}-i{{{\gamma }'}_{u}})-({{\omega }_{x(y)}}-i{{{\gamma }'}_{x(y)}})$,
$\Delta _{p}^{x(y)}=({{\omega }_{x(y)}}-i{{{\gamma }'}_{x(y)}})-\omega _{p}^{x(y)}$ in which ${{{\gamma }'}_{u}}$ and ${{{\gamma }'}_{x(y)}}$ are, respectively, the dephasing rates of biexciton and x (y) exciton states \cite{40}, and
$\bm{\hat{a}}_{p}^{y}{{,}_{{}}}\bm{\hat{a}}_{p}^{x}$ are the
principal plasmon modes annihilation operators. Also,
$\bm{\hat{f}}(\bm{{r}'},{{\omega }_{q}})$ and
${{\bm{\hat{f}}}^{\dagger }}(\bm{{r}'},{{\omega }_{q}})$ denote,
respectively, the bosonic annihilation and creation field operators
for the elementary excitations of the metal nanoparticle for the
first plasmonic reservoir which is coupled only to the
$x(y)\leftrightarrow g$ transition. Moreover\texttt{\emph{,}} the
reservoir operators $\bm{\hat{f}}(\bm{{r}''},{{\omega }_{k}})$ and
${{\bm{\hat{f}}}^{\dagger }}(\bm{{r}''},{{\omega }_{k}})$ correspond
to the second plasmonic reservoir which is coupled only to the
$x(y)\leftrightarrow u$ transition. The coupling strength of the
x-polarized (y-polarized) principal plasmon mode to the environment
is ${{\Omega }_{xm}}({{\Omega }_{ym}})$. The environment is modeled
as a collection of bosonic harmonic oscillators characterized by the
operators $\bm{\hat{f}}({{\omega }_{m}})$ and
${{\bm{\hat{f}}}^{\dagger }}({{\omega }_{m}})$.
Moreover\texttt{\emph{,}} $g_{ex}^{i},g_{bx}^{i}(i=x,y)$ represent
the coupling strengths of the interaction between the principal
plasmon modes and the exciton photon and biexciton photon,
respectively. The frequency of the emitted photons from the
principal plasmon modes is represented by ${{\omega }_{m}}$. Also,
${{\omega }_{u}}{{,}_{{}}}{{\omega }_{x}}_{{}},$ and ${{\omega
}_{y}}$ are the frequencies of the biexciton and excitons states,
respectively. ${{d}_{ij}}$ is the transition dipole moment between
$i$ and $j$ levels and, without lack of generality, we assume the
dipole to be oriented along the $z$ axis. The imaginary part of the
dielectric constant of the MNP is represented as ${{\varepsilon
}_{I}}(\bm{r},\omega )$ and $\bm{G}({{\bm{r}}_{d}},\bm{r},\omega )$
is the field dyadic Green's function \cite{51}.
\par
It is well known that when the environment correlation functions
decay over time scales much shorter than the fastest time scale of
the free system evolution, memory effects can be neglected and one
may approximate the dynamics of the system by the Markovian one. The
MNP acts as a highly structured reservoir for the QD to which it is
strongly coupled. Thus the dynamics of such a hybrid system is
non-Markovian [34]. The QD-MNP dissipative dynamics is determined by
the rapidly structure changing of the reservoir through the
displacement of the QD. When the QD moves away from the MNP, the
reservoir structure becomes smooth, and thus the system experiences
a Markovian dynamics. A comparison between the LSP lifetime, $\tau
\approx 2\pi /{{\kappa }_{sp}}$, and the QD decay rate, which is
determined by the Fermi golden rule, determines the nature of the
system evolution (Markovian or non-Markovian). For example, if a QD
is located at a distance of 10nm from a 7nm radius MNP, the decay
rate of the QD is about 0.5 meV and the decay rate of the LSPs is
about 55 meV (see Fig. (3)). Thus, the characteristic time of the
reservoir is at least two orders of magnitude smaller than that of
the QD. The Markovian behavior becomes more significant when the QD
is placed at a distance of 14nm from the MNP with a 14 nm radius. In
this case, the characteristic time of the reservoir is at least
three orders of magnitude smaller than that of the QD.

\subsection{State of the system}

We assume that the QD is initially prepared in the biexciton state
via electrical pumping, while the other excitations and principal
plasmon modes are in their ground states. Therefore, the state of
the whole system at time $t$ can be written as:
\begin{eqnarray}\label{eq:3}
     \left| \psi (t) \right\rangle &=&{{\left| \psi (t) \right\rangle }_{ind}}+
     {{\left| \psi (t) \right\rangle }_{Plas.scattering}}\nonumber\\&+&{{\left| \psi (t) \right\rangle }_{QDscattering}}.
\end{eqnarray}
The state of the total system is composed of three states including
the states of induced plasmonic modes by the QD, $\left| \psi (t)
\right\rangle ={{\left| \psi (t) \right\rangle }_{ind}}$, the states
of plasmonic modes scattered by the plasmonic reservoirs, ${{\left|
\psi (t) \right\rangle }_{Plas.scattering}}$, and the states of the
scattered QD by the QD-reservoir, ${{\left| \psi (t) \right\rangle
}_{QDscattering}}$.  The first term in Eq.(\ref{eq:3}) is written
as:
    \begin{align}\label{eq:4}
    & {{\left| \psi (t) \right\rangle }_{ind}}={{C}_{1}}(t)\left| u,0,0 \right\rangle {{\left| 0 \right\rangle }_{x}}{{\left| 0 \right\rangle }_{y}}&& \\
    & +C_{2}^{x}(t)\left| x,1,0 \right\rangle {{\left| 0 \right\rangle }_{x}}{{\left| 0 \right\rangle }_{y}} +C_{2}^{y}(t)\left| y,0,1 \right\rangle {{\left| 0 \right\rangle }_{x}}{{\left| 0 \right\rangle }_{y}}&&\nonumber\\
    & +C_{3}^{x}(t)\left| g,2,0 \right\rangle {{\left| 0 \right\rangle }_{x}}{{\left| 0 \right\rangle }_{y}}+C_{3}^{y}(t)\left| g,0,2 \right\rangle {{\left| 0 \right\rangle }_{x}}{{\left| 0 \right\rangle }_{y}}\texttt{\emph{,} } \nonumber&&
    \end{align}
in which the first term indicates that the QD is in the biexciton
state with no principal plasmon mode excitation; the second and
third terms correspond to the QD in the exciton states with one
induced principal plasmon mode excitation in the x-polarized and in
the y-polarized exciton states, respectively; the fourth and fifth
terms describe the situation where the QD is in the ground state and
the two induced principal plasmon mode excitations in the x and y
polarizations are created. The second term in Eq.(\ref{eq:3}) can be
written as:
    \begin{align}\label{eq:5}
    & {{\left| \psi (t) \right\rangle }_{Plas.scattering}}=\\
    &\int{d{{\omega }_{m}}C_{4m}^{x}(t)\left| x,0,0 \right\rangle {{\left| {{1}_{m}} \right\rangle }_{x}}{{\left| 0 \right\rangle }_{y}}} \nonumber\\
    & +\int{d{{\omega }_{m}}C_{4m}^{y}(t)\left| y,0,0 \right\rangle {{\left| 0 \right\rangle }_{x}}{{\left| {{1}_{m}} \right\rangle }_{y}}} \nonumber\\
    & +\int{d{{\omega }_{m}}C_{5m}^{x}(t)\left| g,1,0 \right\rangle {{\left| {{1}_{m}} \right\rangle }_{x}}{{\left| 0 \right\rangle }_{y}}}\nonumber \\
    & +\int{d{{\omega }_{m}}C_{5m}^{y}(t)\left| g,0,1 \right\rangle {{\left| 0 \right\rangle }_{x}}{{\left| {{1}_{m}} \right\rangle }_{y}}} \nonumber\\
    & +\int{d{{\omega }_{m}}\int{d{{\omega }_{n}}}C_{mn}^{x}(t)\left| g,0,0 \right\rangle {{\left| {{1}_{m}},{{1}_{n}} \right\rangle }_{x}}{{\left| 0 \right\rangle }_{y}}}\nonumber \\
    & +\int{d{{\omega }_{m}}\int{d{{\omega }_{n}}}C_{mn}^{y}(t)\left| g,0,0 \right\rangle {{\left| 0 \right\rangle }_{x}}{{\left| {{1}_{m}},{{1}_{n}} \right\rangle }_{y}}}\texttt{\emph{,}} \nonumber
    \end{align}
where different terms describe the possible ways through which the
induced principal plasmon modes are scattered to the plasmonic
reservoir modes. The last term in Eq.(\ref{eq:3}) corresponds to the
decay of QD and is given by:
    \begin{align}\label{eq:6}
    &{{\left| \psi (t) \right\rangle }_{QD scattering}}=\\
    &  \int{{{d}^{3}}\bm{{r}''}\int_{0}^{\infty }{d{{\omega }_{k}}C_{6}^{x}(\bm{{r}''},{{\omega }_{k}},t)\left| x,0,0 \right\rangle }} \left| 1(\bm{{r}''},{{\omega }_{k}}) \right\rangle\nonumber \\
    &   +\int{{{d}^{3}}\bm{{r}''}\int_{0}^{\infty }{d{{\omega }_{k}}C_{6}^{y}(\bm{{r}''},{{\omega }_{k}},t)}} \nonumber \\
    &   \left| y,0,0 \right\rangle \left| 1(\bm{{r}''},{{\omega }_{k}}) \right\rangle +\int{{{d}^{3}}\bm{{r}''}\int_{0}^{\infty }{d{{\omega }_{k}}}} \nonumber \\
    &   \int{{{d}^{3}}\bm{{r}'}\int_{0}^{\infty }{d{{\omega }_{q}}}}(C_{7}^{x}(\bm{{r}''},\bm{{r}'},{{\omega }_{k}},{{\omega }_{q}},t)\left| g,0,0 \right\rangle  \nonumber \\
    &   \left| 1(\bm{{r}''},{{\omega }_{k}})\texttt{\emph{,}}1(\bm{{r}'},{{\omega }_{q}}) \right\rangle )+\int{{{d}^{3}}\bm{{r}''}\int_{0}^{\infty }{d{{\omega }_{k}}}} \nonumber \\
    &   \int{{{d}^{3}}\bm{{r}'}\int_{0}^{\infty }{d{{\omega }_{q}}}}(C_{7}^{y}(\bm{{r}''},\bm{{r}'},{{\omega }_{k}},{{\omega }_{q}},t)\left| g,0,0 \right\rangle  \nonumber \\
    &   \left| 1(\bm{{r}''},{{\omega }_{k}})\texttt{\emph{,}}1(\bm{{r}'},{{\omega }_{q}}) \right\rangle ) \nonumber \\
    &    +\int{{{d}^{3}}\bm{{r}''}\int_{0}^{\infty }{d{{\omega }_{k}}C_{8}^{x}(\bm{{r}''},{{\omega }_{k}},t)\left| g,1 \right\rangle }}\left| 1(\bm{{r}''},{{\omega }_{k}}) \right\rangle  \nonumber \\
    &    +\int{{{d}^{3}}\bm{{r}''}\int_{0}^{\infty }{d{{\omega }_{k}}C_{8}^{y}(\bm{{r}''},{{\omega }_{k}},t)\left| g,0 \right\rangle }}\left| 1(\bm{{r}''},{{\omega }_{k}}) \right\rangle  \nonumber \\
    &  +\int{{d}^{3}}\bm{{r}''}\int_{0}^{\infty }{d{{\omega }_{q}}C_{9}^{x}(\bm{{r}'},{{\omega }_{k}},t)\left| g,1 \right\rangle }\left| 1(\bm{{r}''},{{\omega }_{k}}) \right\rangle  \nonumber \\
    &    +\int{{{d}^{3}}\bm{{r}'}\int_{0}^{\infty }{d{{\omega }_{q}}C_{9}^{y}(\bm{{r}'},{{\omega }_{q}},t)\left| g,1 \right\rangle }}\left| 1(\bm{{r}'},{{\omega }_{k}}) \right\rangle\texttt{\emph{,}}  \nonumber
    \end{align}
where the different terms describe the spontaneous decay of the
exciton and biexciton states due to the plasmonic reservoir modes.
As a consequence of the presence of the MNP the local density of
state (LDOS)
of the environment will be increased (Purcell effect).\\
\par Using the time-dependent Schr$\ddot{o}$dinger equation, we
arrive at the following equations of motion for the probability
amplitudes:
    \begin{subequations}
        \begin{align}\label{eq:7a}
        & {{{\dot{C}}}_{1}}(t)=-ig_{bx}^{x}C_{2}^{x}(t){{e}^{-i({{\omega }_{ux}}-\omega _{p}^{x})t}}&&\nonumber\\
        &-ig_{bx}^{y} C_{2}^{y}(t){{e}^{-i({{\omega }_{u}}_{y}-\omega _{p}^{y})t}}-{{\gamma }_{bx}}{{C}_{1}}(t)\texttt{\emph{,}} \\
        & \dot{C}_{2}^{\alpha }(t)=-ig_{2}^{\alpha }{{C}_{1}}(t){{e}^{-i({{\omega }_{u}}-{{\omega }_{x\alpha }}-\omega _{p}^{\alpha })t}}\nonumber\\
        &-ig_{ex}^{\alpha } \sqrt{2}C_{3}^{\alpha }(t){{e}^{i\Delta _{p}^{\alpha }t}}-{{\gamma }_{bx}}{{C}_{2}}(t)\nonumber\\
        &-{{\kappa }_{\alpha }}C_{2}^{\alpha }(t)\texttt{\emph{,}} \\
        &  \dot{C}_{3}^{\alpha }(t)=-ig_{ex}^{\alpha }\sqrt{2}C_{2}^{\alpha }(t){{e}^{-i\Delta _{p}^{\alpha }t}}\nonumber\\
        &-2{{\kappa }_{\alpha }} C_{3}^{\alpha }(t)\texttt{\emph{,}}\\
        & \dot{C}_{4m}^{\alpha }(t)=-ig_{ex}^{\alpha }C_{5m}^{\alpha }(t){{e}^{i\Delta _{p}^{\alpha }t}}\nonumber\\
        &-i\Omega _{\alpha m}^{*}C_{2}^{\alpha }(t) {{e}^{-i(\omega _{p}^{\alpha }-{{\omega }_{m}})t}}-{{\gamma }_{ex}}C_{4m}^{\alpha }(t)\texttt{\emph{,}} \\
        & {{{\dot{C}}}_{5m}}(t)=-ig_{ex}^{\alpha }C_{4m}^{\alpha }(t){{e}^{-i\Delta _{p}^{\alpha }t}}\nonumber\\
        &-i\Omega _{\alpha m}^{*}\sqrt{2}C_{3}^{\alpha }(t){{e}^{-i(\omega _{p}^{\alpha }-{{\omega }_{m}})t}}-{{\kappa }_{\alpha }}C_{5m}^{\alpha }(t)\texttt{\emph{,}} \\
        &   \dot{C}_{mn}^{\alpha }(t)=-i\Omega _{\alpha n}^{*}\sqrt{2}C_{5m}^{\alpha }(t){{e}^{-i(\omega _{p}^{\alpha }-{{\omega }_{n}})t}}.\label{eq:7f}
        \end{align}
    \end{subequations}
In these equations, $\alpha =x$ or $y$, ${{\kappa }_{\alpha }}=\pi
|{{\Omega }_{\alpha m}}{{|}^{2}}$ is the spectral half  width of the
plasmonic peak, ${{\gamma }_{ex}}=\frac{{{\omega
}_{x}}^{2}}{{{c}^{2}}\hbar {{\varepsilon
}_{0}}}{{\bm{d}}_{gx}}.\operatorname{Im}\bm{G}({{r}_{d}},{{r}_{d}},{{\omega
}_{x}}).{{\bm{d}}_{gx}}$, and ${{\gamma }_{bx}}\simeq 2{{\gamma
}_{ex}}$ denote, respectively, the decay rates of the exciton and
biexciton states including both radiative and nonradiative
broadening. Moreover\texttt{\emph{,}} without loss of generality, we
supposed that $|{{\bm{d}}_{gx}}|=|{{\bm{d}}_{ux}}|$.\\
Our purpose is to investigate the influence of the MNP plasmons on
the formation of entangled photon pairs from the QD. Therefore, the
prepared QD in the biexciton state should experience a cascade
transition to the ground state with the emission of two photons. For
this reason, we are interested in finding the field state in the
long time limit\texttt{\emph{,}} i.e.\texttt{\emph{,}} $t\gg
{{\gamma }_{ex}}^{-1},{{\gamma }_{bx}}^{-1},{{\kappa }^{-1}}$. By
applying the Laplace transform method to solve the set of
Eqs.(\ref{eq:7a}-\ref{eq:7f})\texttt{\emph{,}} the probability
amplitudes for the emission of two photons from the QD in the
vicinity of the MNP in the long time limit can be obtained as:
    \begin{subequations}
        \begin{align}   \label{eq:8a}
        & c_{mn}^{x}(\infty )=\frac{g_{bx}^{x}\Omega _{xm}^{*}{{F}_{y}}({{\omega }_{m}},{{\omega }_{n}})}{D({{\omega }_{m}},{{\omega }_{n}})}\times  &&\\
        \label{eq:8b} & \frac{g_{ex}^{x}\Omega _{xn}^{*}({{\omega }_{m}}+3{{\omega }_{n}}-2{{\omega }_{x}}-2\omega _{p}^{x}+2i{{\kappa }_{x}}+2i{{\gamma }_{ex}})}{({{\omega }_{n}}-{{\omega }_{x}}+i{{\gamma }_{ex}})({{\omega }_{n}}-\omega _{p}^{x}+i{{\kappa }_{x}})-{{(g_{ex}^{x})}^{2}}}\texttt{\emph{,}}
        \nonumber\\
        & c_{mn}^{y}(\infty )=\frac{g_{bx}^{y}\Omega _{ym}^{*}{{F}_{x}}({{\omega }_{m}},{{\omega }_{n}})}{D({{\omega }_{m}},{{\omega }_{n}})}\times  \\
        & \frac{g_{ex}^{y}\Omega _{yn}^{*}({{\omega }_{m}}+3{{\omega }_{n}}-2{{\omega }_{y}}-2\omega _{p}^{y}+2i{{\kappa }_{y}}+2i{{\gamma }_{ex}})}{({{\omega }_{n}}-{{\omega }_{y}}+i{{\gamma }_{ex}})({{\omega }_{n}}-\omega _{p}^{y}+i{{\kappa }_{y}})-{{(g_{ex}^{y})}^{2}}}\texttt{\emph{,}} \nonumber
        \end{align}
    \end{subequations}
where the functions $D(\omega_{m}, \omega_{n})$ and
$F_{\alpha}(\omega_{m}, \omega_{n})$,
    $\alpha = x, y$, are defined by:
    \begin{subequations}
        \begin{align}\label{eq:9}
        & D({{\omega }_{m}},{{\omega }_{n}})=({{\omega }_{m}}+{{\omega }_{n}}-{{\omega }_{u}}+i{{\gamma }_{bx}})\times&& \nonumber\\
        & {{F}_{x}}({{\omega }_{m}},{{\omega }_{n}}){{F}_{y}}({{\omega }_{m}},{{\omega }_{n}})+{{(g_{bx}^{x})}^{2}} {{F}_{y}}({{\omega }_{m}},{{\omega }_{n}})\times \nonumber\\
        &({{\omega }_{m}}+{{\omega }_{n}}-2\omega _{p}^{x}+2i{{\kappa }_{x}})+
        {{(g_{bx}^{y})}^{2}}{{F}_{x}}({{\omega }_{m}},{{\omega }_{n}})\nonumber\\
        &\times({{\omega }_{m}}+{{\omega }_{n}}-2\omega _{p}^{y}+2i{{\kappa }_{y}})\texttt{\emph{,}} \\
        & {{F}_{\alpha }}({{\omega }_{m}},{{\omega }_{n}})=2{{(g_{ex}^{\alpha })}^{2}}-({{\omega }_{m}}+{{\omega }_{n}}-{{\omega }_{\alpha }}- \nonumber\\
        & \omega _{p}^{\alpha }+i{{\kappa }_{x}}+i{{\gamma }_{ex}})({{\omega }_{m}}+{{\omega }_{n}}-2\omega _{p}^{\alpha }+2i{{\kappa }_{x}}).
        \end{align}
    \end{subequations}
Furthermore, ${{\omega }_{n}}$ and ${{\omega }_{m}}$ denote the
biexciton-exciton and exciton- ground state transition frequencies,
respectively.
\section{SPECTRUM AND POLARIZATION ENTANGLEMENT OF THE GENERATED PHOTON PAIRS}
Having determined the state of the whole system at the long time
limit, we are now in a position to investigate the physical
properties of the system. By using Eqs. (4a) and
(4b)\texttt{\emph{,}} one can determine the spectral
functions\texttt{\emph{,}} i.e.\texttt{\emph{,}} the joint
probability distributions of the emitted x-polarized and y-polarized
photon pairs, defined by ${{S}_{\alpha }}({{\omega }_{m}},{{\omega
}_{n}})=|c_{mn}^{\alpha }(\infty ){{|}^{2}}(\alpha =x,y)$. The
spectrum of the x-polarized (y-polarized) photons coming from the
biexciton-exciton (exciton-ground state) transition is obtained by
integrating ${{S}_{x}}({{\omega }_{m}},{{\omega
}_{n}})[{{S}_{y}}({{\omega }_{m}},{{\omega }_{n}})]$ over ${{\omega
}_{n}}({{\omega }_{m}})$\texttt{\emph{,}} i.e.\texttt{\emph{,}}
    \begin{subequations}
        \begin{align}\label{eq:10}
        & {{S}_{x}}({{\omega }_{m}})=\int_{-\infty }^{\infty }{d{{\omega }_{n}}}|c_{mn}^{x}(\infty ){{|}^{2}}\texttt{\emph{,}}&&\\
        & {{S}_{x}}({{\omega }_{n}})=\int_{-\infty }^{\infty }{d{{\omega }_{m}}}|c_{mn}^{x}(\infty ){{|}^{2}}\texttt{\emph{,}}&&
        \end{align}
    \end{subequations}
    and
    \begin{subequations}
        \begin{align}\label{eq:11}
        & {{S}_{y}}({{\omega }_{m}})=\int_{-\infty }^{\infty }{d{{\omega }_{n}}}|c_{mn}^{y}(\infty ){{|}^{2}}\texttt{\emph{,}}&&\\
        & {{S}_{y}}({{\omega }_{n}})=\int_{-\infty }^{\infty }{d{{\omega }_{m}}}|c_{mn}^{y}(\infty ){{|}^{2}}.&&
        \end{align}
    \end{subequations}

\par In the steady state, the general form of the wave function of
the photon pairs emitted through both x and y intermediate channels
can be written as
    \begin{align}\label{eq:12}
    & \left| \psi (\infty ) \right\rangle =\alpha \left| {{P}_{x}} \right\rangle \left| xx \right\rangle +\beta \left| {{P}_{y}} \right\rangle \left| yy \right\rangle \texttt{\emph{,}} &&\nonumber\\
    & |\alpha {{|}^{2}}+|\beta {{|}^{2}}=1\texttt{\emph{,}}
    \end{align}
where $\alpha$ and $\beta$ are the probability amplitudes for the
two possible decay channels, and $\left| {{P}_{x}} \right\rangle $
and $\left| {{P}_{y}} \right\rangle $ represent, respectively, the
coordinate parts of the two-photon wave packets for the x- and y-
polarizations. Moreover, $\left| xx \right\rangle $ and $\left| yy
\right\rangle $ are the corresponding polarization parts of the wave
function. By tracing out the coordinate part of the two-photon wave
packet, the reduced density matrix containing all information about
the two-photon polarization quantum state is obtained as:
    \begin{align}\label{eq:13}
    & \hat{\rho }=\left( \begin{matrix}
    |\alpha {{|}^{2}} & 0 & 0 & \gamma   \\
    0 & 0 & 0 & 0  \\
    0 & 0 & 0 & 0  \\
    {{\gamma }^{*}} & 0 & 0 & |\beta {{|}^{2}}  \\
    \end{matrix} \right)\texttt{\emph{,}}
    & \gamma =\alpha {{\beta }^{*}}\left\langle {{P}_{y}}|{{P}_{x}} \right\rangle.
    \end{align}
\par Each photon has two polarization degrees of freedom. Thus, a
system consisting of two photons is a two-qubit system. A number of
criteria have been proposed to establish whether or not a given
density matrix of a system is separable. As far as a two-qubit
system is concerned, concurrence is a suitable quantitative measure
to identify the polarization entanglement \cite{52}, which is
defined as $C(\rho )=\max \{0,{{\lambda }_{1}}-{{\lambda
}_{2}}-{{\lambda }_{3}}-{{\lambda }_{4}}\}$. In this definition,
${{\lambda }_{1}},...,{{\lambda }_{4}}$ are the square roots of the
eigenvalues in decreasing order of magnitude of the matrix
$\bm{\hat{R}}=\bm{\hat{\rho }\hat{\tilde{\rho }}}$ where
$\hat{\tilde{\rho }}=({{\hat{\sigma }}_{y}}\otimes {{\hat{\sigma
}}_{y}}){{\hat{\rho }}^{*}}({{\hat{\sigma }}_{y}}\otimes
{{\hat{\sigma }}_{y}})$. The range of concurrence is from 0 for
separable states, to 1 for maximally entangled pure states. For the
system under consideration the concurrence is given by $C(\rho
)=2|\gamma |$. Evidently, if the states $\left| {{P}_{x}}
\right\rangle $ and $\left| {{P}_{y}} \right\rangle $ are orthogonal
to each other the two transition channels are distinguishable and
consequently, no polarization entanglement will arise. On the other
hand , if $\alpha=\beta$ and $\left| {{P}_{x}} \right\rangle $ is
parallel to $\left| {{P}_{y}} \right\rangle $ then $|\gamma|=1/2$
which corresponds to maximally-polarization-entangled state.\\
\par The polarized entangled photon pairs can be categorized by their
energies; one pair resulting from the biexciton-exciton transition
with the mean energy of $E=({{\omega }_{x}}+{{\omega
}_{y}})/2-{{\Delta }_{xx}}$, and another one resulting from
exciton-ground state transition with a mean energy of $E=({{\omega
}_{x}}+{{\omega }_{y}})/2$. It is possible to define spectral
windows of a detector to count only one group of the emitted
entangled photons. Mathematically, this procedure can be done by
applying a projection operator on the wave packet, to get the
normalized two-photon wavefunction $\left| {\psi(\infty ) }
\right\rangle =P\left| {\psi(\infty )} \right\rangle /|P\left|
{\psi(\infty )} \right\rangle |$. As the result, the off-diagonal
element of the density matrix (\ref{eq:13})\texttt{\emph{,}} by
which the magnitude of polarization entanglement is determined,
takes the form ${\gamma }'=\alpha {{\beta }^{*}}\left\langle
{{P}_{y}}|P|{{P}_{x}} \right\rangle /|P\left| {\psi(\infty ) }
\right\rangle {{|}^{2}}$.\\
\par In the system under consideration, the wave function of the
photon pairs emitted through both x and y intermediate channels can
be written as:
    \begin{align}\label{eq:14}
    & \left| \psi (\infty ) \right\rangle = &&\\
    & \int{d{{\omega }_{m}}\int{d{{\omega }_{n}}}C_{mn}^{x}(\infty )\left| g,0,0 \right\rangle {{\left| {{1}_{m}},{{1}_{n}} \right\rangle }_{x}}{{\left| 0 \right\rangle }_{y}}+}\nonumber \\
    & \int{d{{\omega }_{m}}\int{d{{\omega }_{n}}}C_{mn}^{y}(\infty )\left| g,0,0 \right\rangle {{\left| 0 \right\rangle }_{x}}{{\left| {{1}_{m}},{{1}_{n}} \right\rangle }_{y}}}\texttt{\emph{,}} \nonumber
    \end{align}
where the probability amplitudes $C_{mn}^{i}(i=x,y)$ are given in
Eqs. (\ref{eq:8a}) and (\ref{eq:8b}). Thus the off-diagonal element
of the reduced density matrix associated with the two-photon
emission via the scattering of two principal plasmon modes to the
environment is obtained as
    \begin{align}\label{eq:15}
    {{\gamma }^{'}}=\frac{P}{T+H}\texttt{\emph{ ,}}
    \end{align}
    where
    \begin{subequations}
        \begin{align}\label{eq:15b}
        & P=\int{\int{d{{\omega }_{m}}d{{\omega }_{n}}C_{mn}^{{{x}^{*}}}(\infty ).C_{mn}^{y}(\infty )W\texttt{\emph{,}}}}&&  \\
        & T=\int{\int{d{{\omega }_{m}}d{{\omega }_{n}}|C_{mn}^{x}(\infty ){{|}^{2}}W}}\texttt{\emph{,}}&& \\
        & H=\int{\int{d{{\omega }_{m}}d{{\omega }_{n}}|C_{mn}^{y}(\infty ){{|}^{2}}W}}.
        \end{align}
    \end{subequations}
with W being the spectral window function whose value is zero or one
\cite{22,24}.
    \section{RESULTS AND DISCUSSIONS}
In this section, we present and discuss the results of the numerical
calculations to investigate the polarization entanglement of the
photon pairs which are emitted in the system under consideration.
Throughout the calculations, we use atomic units $($ $\hbar =1$,
$4\pi {{\varepsilon }_{0}}=1$, $c=137$, and $e=1)$.
    \subsection{LDOS of the system}
We consider a QD as the emitter with the dipole moment of $|d|=0.5 e
nm$ located at the distance $h$ from the surface of a spherical
silver MNP of radius $R$ with the permittivity function given by the
generalized Drude model, $\varepsilon (\omega )=\varepsilon (\infty
)-\frac{\omega _{p}^{2}}{{{\omega }^{2}}-i\gamma \omega }$, with
$\varepsilon (\infty )=6$, the Landau damping constant $\gamma =51
meV$, and the plasma frequency ${{\omega }_{p}}=7.9 eV$
\cite{53,54}.
\par
Following the experiments on the decay rate of a III/V self-assembled quantum dot \cite{20-5}, the pure dephasing rate is estimated to be ${{{\gamma }'}_{x(y)}}\simeq 1\mu eV$ (the dephasing lifetime for biexciton: 405 ps, exciton: 771 ps). Our numerical results show that the pure dephasing in the system under consideartion has a negligible effect on the entanglement between photon pairs generated from the biexciton cascade transition even by increasing the rate of pure dephasing up to 0.1meV. In other words, the plasmon coupling dominates the decay of the biexciton cascade transition.
    \begin{figure}
        \includegraphics[width=\linewidth]{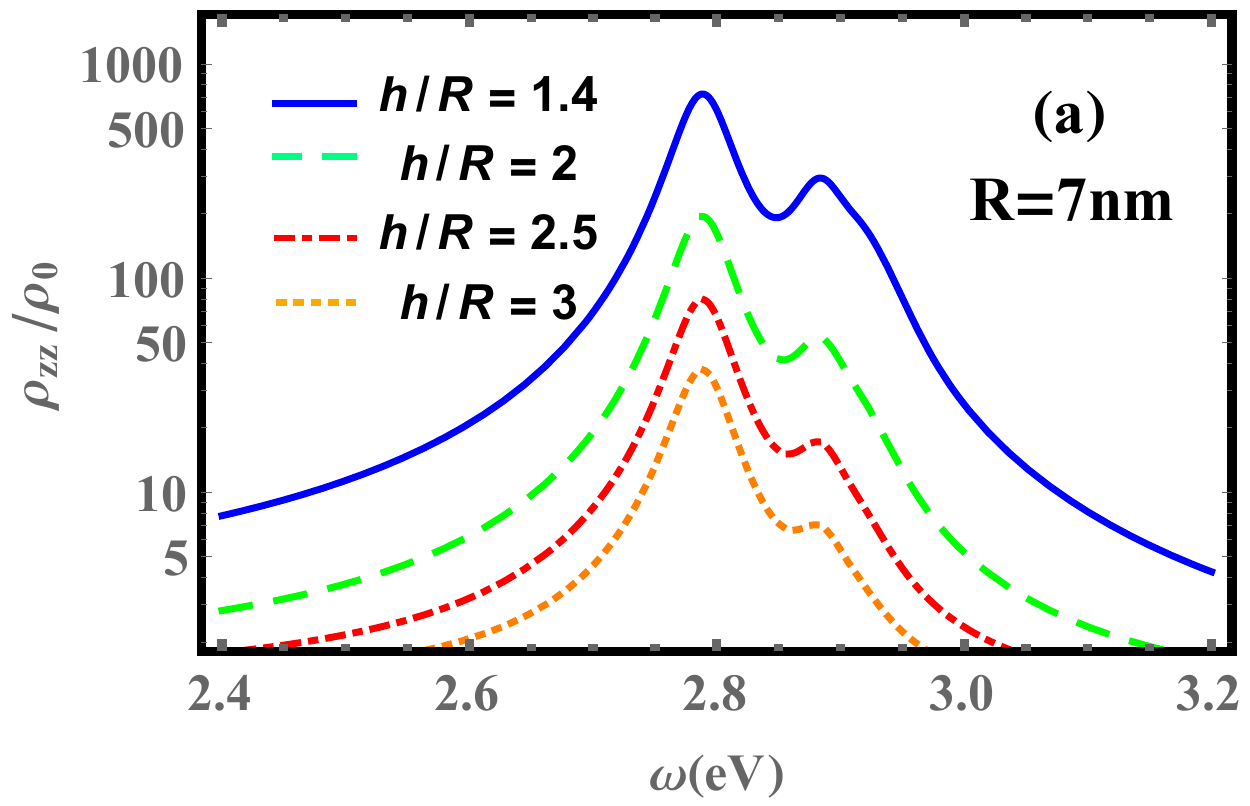}[t]
    \end{figure}
    \begin{figure}
        \includegraphics[width=\linewidth]{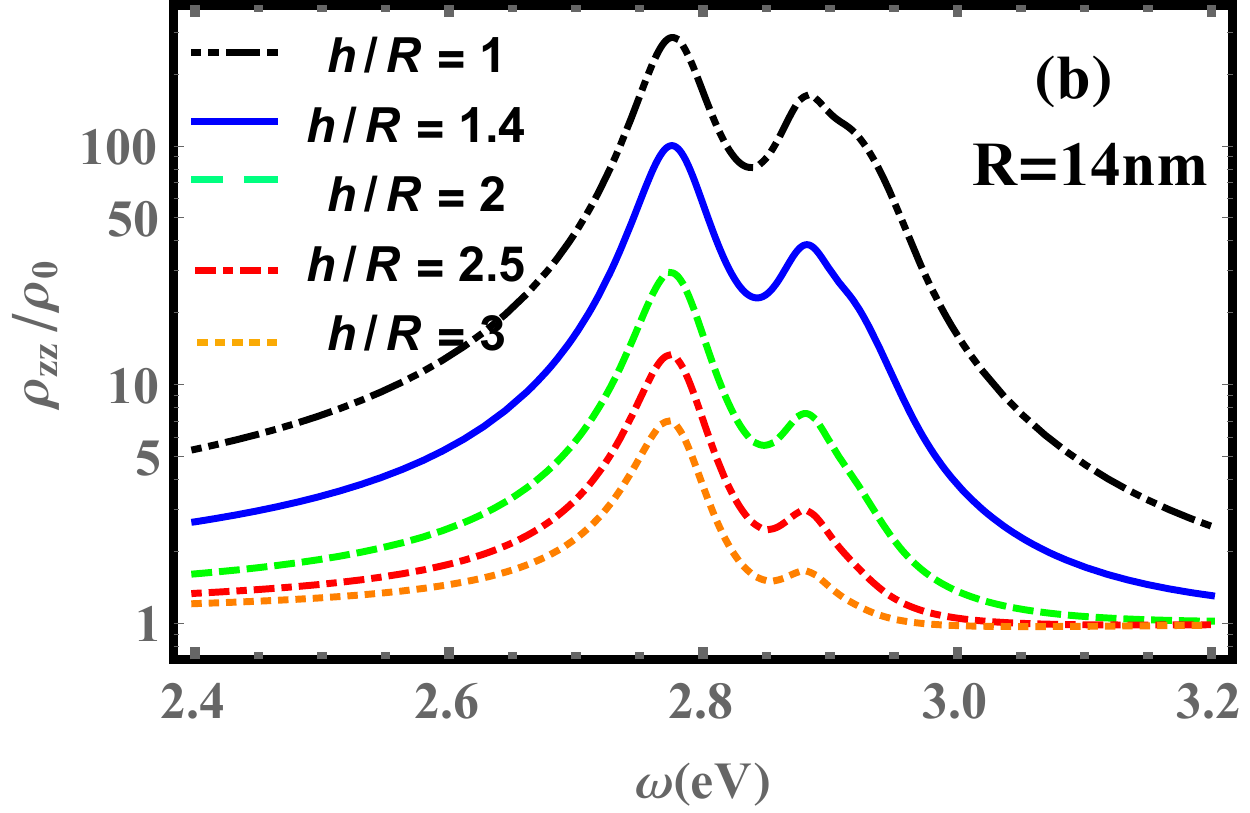}
        \caption{(Color online) Scaled LDOS, ${{\rho }_{zz}}/{{\rho }_{0}}$, as a function of
        frequency $\omega$ for different distances of QD from the surface of the MNP and for
        two values of the MNP radius: (a) $R=7$ nm and (b) $R=14$ nm. For the defined parameters,
        the LDOS at dipole plasmon mode is more significant and larger than that at higher-order plasmon mode.}
        \label{fig:schematic diagram}
    \end{figure}
\par In Fig. (2) we have plotted the scaled LDOS, ${{\rho
}_{zz}}/{{\rho
}_{0}}=\operatorname{Im}[{{G}_{zz}}({{r}_{d}},{{r}_{d}},\omega
)]/{{\rho }_{0}},$ with ${{\rho }_{0}}=k_{1}/6\pi $ being the free
space DOS, versus frequency $\omega $ for different values of the
QD-MNP distance and for two values of the MNP radius. Here,
    \begin{align}\label{eq:G2}
    &\operatorname{Im}[{{G}_{zz}}({{r}_{d}},{{r}_{d},\omega)]}=\dfrac{k_{1}}{4 \pi} \operatorname{Re}\sum\limits_{n=1}^{\infty }{(2n+1)}n(n+1)\nonumber\\
    &\times{{R}^{V}}{{[\frac{h_{n}^{(1)}({{k}_{1}}r)}{{{k}_{1}}r}]}^{2}}\texttt{\emph{,}}
    \end{align}
where $h^{(1)}_{n}$ is the spherical Hankel function of the first
kind and the coefficient $R^{V}$ is given by \cite{53}
    \begin{widetext}
    \begin{eqnarray} {{R}^{V}}=\frac{k_{1}^{2}{{j}_{n}}({{k}_{1}}R)\frac{d({{k}_{2}}r{{j}_{n}}({{k}_{2}}r))}{d({{k}_{2}}r)}{{|}_{r=R}}-k_{2}^{2}{{j}_{n}}({{k}_{2}}R)\frac{d({{k}_{1}}r{{j}_{n}}({{k}_{1}}r))}{d({{k}_{1}}r)}{{|}_{r=R}}} {k_{2}^{2}{{j}_{n}}({{k}_{2}}R)
\frac{d({{k}_{1}}rh_{n}^{(1)}({{k}_{1}}r))}{d({{k}_{1}}r)}{{|}_{r=R}}-k_{1}^{2}h_{n}^{(1)}({{k}_{1}}R)\frac{d({{k}_{2}}r{{j}_{n}}({{k}_{2}}r))}{d({{k}_{2}}r)}{{|}_{r=R}}}\texttt{\emph{,}}
    \end{eqnarray}
    \end{widetext}
with $\bm{{{k}_{1}}}=\frac{\omega }{c}\sqrt{{{\varepsilon }_{b}}}$
and $\bm{{{k}_{2}}}=\frac{\omega }{c}\sqrt{\varepsilon
(\omega,\bm{r} )}$.
\par In Fig. (2a) the spherical MNP is assumed to
have a radius $R=7$ nm and the MNP-QD distance $h$ is set in the
range of 10 to 16 nm. Also, Fig. (2b) represents the LDOS of a $14$
nm radius spherical MNP when the MNP-QD distance $h$ is set in the
range of 16 to 24 nm. Three main results can be identified from Fig.
(2). First, the metal nanosphere is a highly structured reservoir
(instead of single Lorentzian) and the picks which indicate the
response of the system, are determined by the poles of the dyadic
Green's function. The poles occur at two principal modes; the peak
at the dipole plasmon mode (left peaks in Figs. (2a) and (2b)) is in
lower energy state compared to the higher-order plasmon modes (right
peaks in the Fig. (2a) and (2b)). Second, the positions of the peaks
change slightly as the radius of MNP changes. As the third result,
the structure of the reservoir changes considerably by varying the
QD-MNP distance $h$. The decrease of the LDOS at higher-order
plasmon mode is faster than  its decrease at the dipole mode.
    \subsection{The coupling strength }
\par To reach the strong-coupling regime among the QD and MNP in this
hybrid system, one requires to put the QD in the near field of the
MNP so that the QD-MNP coupling strength $g$ could be larger than
any dissipation decay rate in the system. In the system under
consideration, the strong coupling between the QD exciton and
higher-order plasmon modes which manifests itself as the vacuum Rabi
splitting in the spectrum emitted by the QD, is more  significant
whenever the QD-MNP separation is less  than half of the radius of
the MNP  \cite{53}. Nevertheless, the weak coupling regime has its
own interest for effective single-photon or two-photon generation.As
stated before, to ensure the Markovian behavior of the system under
consideration the QD should be located away from the MNP. In this
case, the significant response of the MNP is through the dipole
modes and the hybrid system enters the weak-coupling regime. The
MNP-QD coupling strength in the weak-coupling
regime\texttt{\emph{,}} i.e.\texttt{\emph{,}} in the limit of
$\kappa \gg g_{ex,bx}^{x,y},{{\gamma }_{ex,bx}}$  is given by
\cite{55}:
    \begin{align}\label{eq:16}
    g=\frac{1}{2}\sqrt{{{\gamma }^{(0)}_{(ex,bx)0}}\kappa {{\rho }_{zz}}/{{\rho }_{0}}}\texttt{\emph{,}}
    \end{align}
where ${{\gamma }^{(0)}_{(ex,bx)}}$ is the decay rate of the QD in
the free space and $\kappa =\gamma +{{\gamma }_{r}}$ is the decay
rate of the plasmonic modes including both ohmic losses in the metal
($\gamma $) and scattering into the free-space modes ( ${{\gamma
}_{r}}$) which can be calculated classically from the Larmor formula
${{\gamma }_{r}}=2\omega _{0}^{4}{{\varepsilon
}_{b}}^{2}{{R}^{3}}/{{c}^{3}}(2{{\varepsilon }_{b}}+1)$, where
${{\varepsilon }_{b}}$ is the dielectric constant outside the MNP
and ${{\omega }_{0}}=({{\omega }_{x}}+{{\omega }_{y}})/2$ \cite{56}.
The decay rate of plasmonic modes ($\kappa $) is frequency
independent because $\gamma $ (the Landau damping rate) in metals is
constant. Furthermore, since ${{\gamma }_{r}}$ depends on the radius
of MNP one has ${{\kappa }_{x}}={{\kappa }_{y}}$. Equation
(\ref{eq:16}) shows also that the coupling strength depends on the
free-space spontaneous emission of the emitter. Therefore, the
dipole moment ${{\gamma }^{(0)}_{(ex,bx)}}$ is directly proportional
to $|d{{|}^{2}}$; the larger dipole moment is, the higher coupling
strength is.
\par The total decay rate of the QD includes the
contributions of both radiative and non-radiative processes with
decay rates ${{\gamma }_{rad}}$ and ${{\gamma }_{nonrad}}$,
respectively. If the QD is placed close to the MNP, it can couple to
both the nonpropagating, quickly decaying evanescent modes which
leads to the energy dissipation through the heating of the MNP, and
to the local surface plasmon modes by which the energy is
transferred to the MNP through the stimulated or spontaneous
emission. As a consequence of significant QD-MNP coupling strength
when the QD is close to the MNP, the decay rate ${{\gamma
}_{nonrad}}$ becomes large. To avoid this quenching effect, one may
displace the QD away from the MNP and place it at an intermediate
region close enough to still couple efficiently with lower-order
plasmon modes. But by almost eliminating the higher-order plasmon
modes we set ${{\gamma }_{nonrad}}=0$ \cite{56,57,58}. This is valid
when $h/R\ge 1.4$ for $R=$ 7 nm  and when $h/R\ge$ 1 for $R=14$ nm.
In Fig. (3) we have plotted the QD-MNP coupling strength (solid
line)\texttt{\emph{,}} given by Eq. (\ref{eq:16})\texttt{\emph{,}}
together with the decay rate of the QD (dashed line) with respect to
the ratio $h/R$ for two different values of the MNP radius. As can
be seen, with increasing the MNP radius the coupling strength
decreases and the QD decays at a slower rate. In both cases, the
QD-MNP coupling strength is greater than the decay rate of the QD.
    \begin{figure}
        \includegraphics[width=\linewidth]{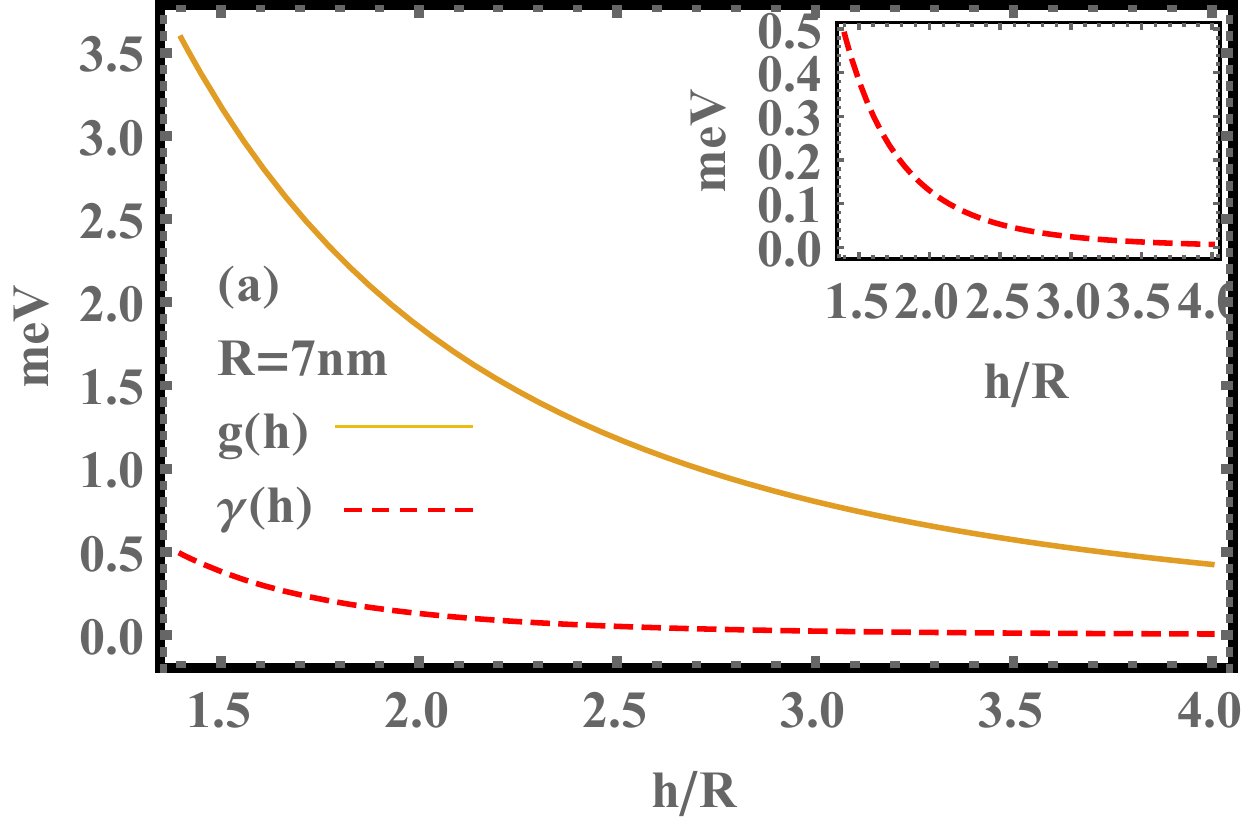}
        \includegraphics[width=\linewidth]{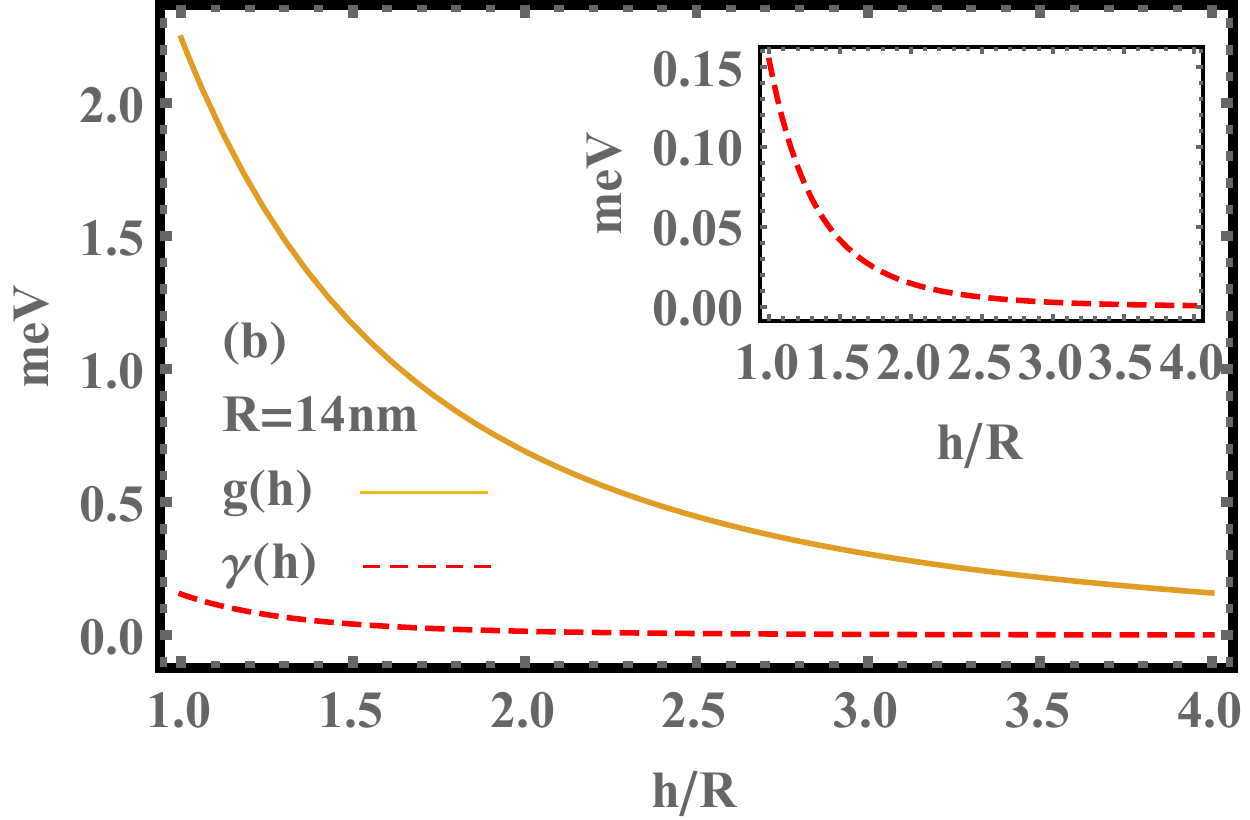}
        \caption{The QD-MNP coupling strength (yellow solid line) and the
        QD decay rate (red dashed line) versus the ratio of the QD-MNP
        distance to the MNP radius $(\frac{h}{R})$ for two different values of the
        MNP radius: (a) $R$= 7 nm and (b) $R$=14nm. The inset represents the decay
         rate dependence on the $\frac{h}{R}$ in a separate panel.}
    \end{figure}
    \subsection{The spectrum of the photon pairs}
    \par In Fig. (4) we have plotted the spectrum of the generated photon pairs with two orthogonal polarizations versus $\omega -{{\omega }_{0}}$ for different values of the QD-MNP separation distance, when the radius of the MNP is $R$=7 nm. These spectra belong to the induced decay of the QD via coupling with the principal plasmon modes and scattering of principal plasmon modes. This figure shows the spectrum of the photons generated in the first transition\texttt{\emph{,}} i.e.\texttt{\emph{,}} the photons resulting from the biexciton-exciton transition for both x-polarization (red dash-dotted line) and y-polarization (blue dashed line) as well as those photons generated via second transition for both x-polarization (orange solid line) and y-polarization (black dotted line).
    \par Two main results can be identified from Figs. 4(a)-(d). Firstly, by increasing the QD-MNP separation distance, not only the spectra of the x-polarized and y-polarized biexciton photons (shown by blue dashed and red dash-dotted lines\texttt{\emph{,}} respectively)\texttt{\emph{,}} but also the spectra of the x-polarized and y-polarized exciton photons (shown by solid orange and black dotted lines\texttt{\emph{,}} respectively) are resolved from each other. Secondly, by increasing the QD-MNP distance the full width at half maximum (FWHM) which is related to the decay rate of each transition is decreased. This is due to the fact that the induced structure of the reservoir by the MNP becomes smooth and small in large distances from the MNP.
    \par As is shown in Figs. 4(c) and 4(d)\texttt{\emph{,}} the central peaks at $\omega -{{\omega }_{0}}=\pm {{\delta }_{x}}/2$ correspond to the pairs of exciton photons (orange and black dotted lines) and the peaks centered at $\omega -{{\omega }_{0}}=-{{\Delta }_{xx}}\pm {{\delta }_{x}}/2$ correspond to the biexciton photons (blue dashed and red dash-dotted lines). In Figs. (4a) and (4b) the values of FWHM are much greater than ${{\delta }_{x}}$, that is the energy difference between the two orthogonal-polarization photons generated through the first and second transitions. Thus, the value of FSS or ${{\delta }_{x}}$ has negligible effects on the spectrum so that the wave functions of the two photons with orthogonal polarizations are overlapped. With increasing the QD-MNP separation distance (see Figs. 4(c) and 4(d)) the FWHM decreases and the influence of FSS on the splitting of the spectral peaks of orthogonal polarizations is observed.
    \begin{figure}
        \includegraphics[width=\linewidth]{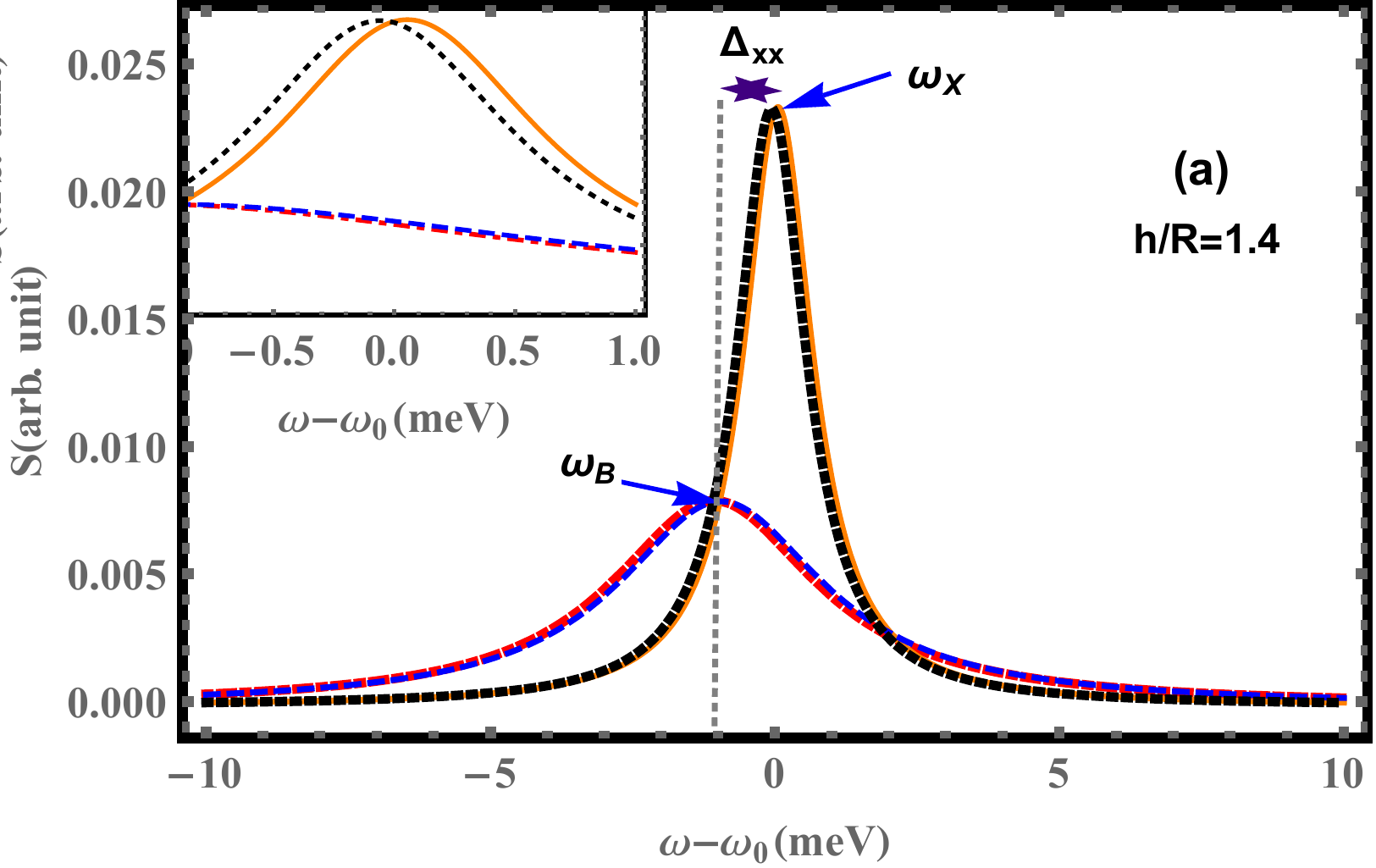}
        \includegraphics[width=\linewidth]{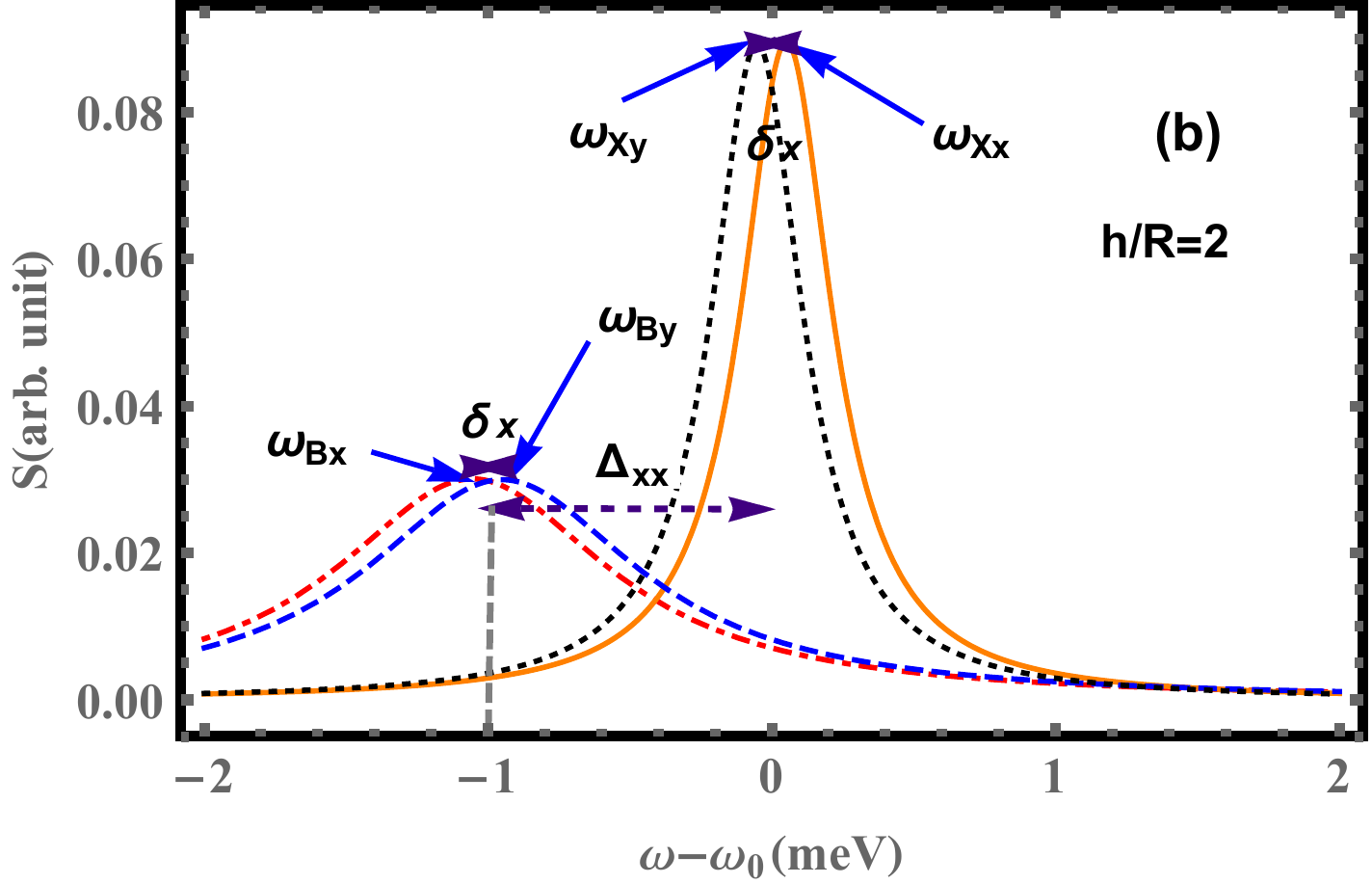}
        \includegraphics[width=\linewidth]{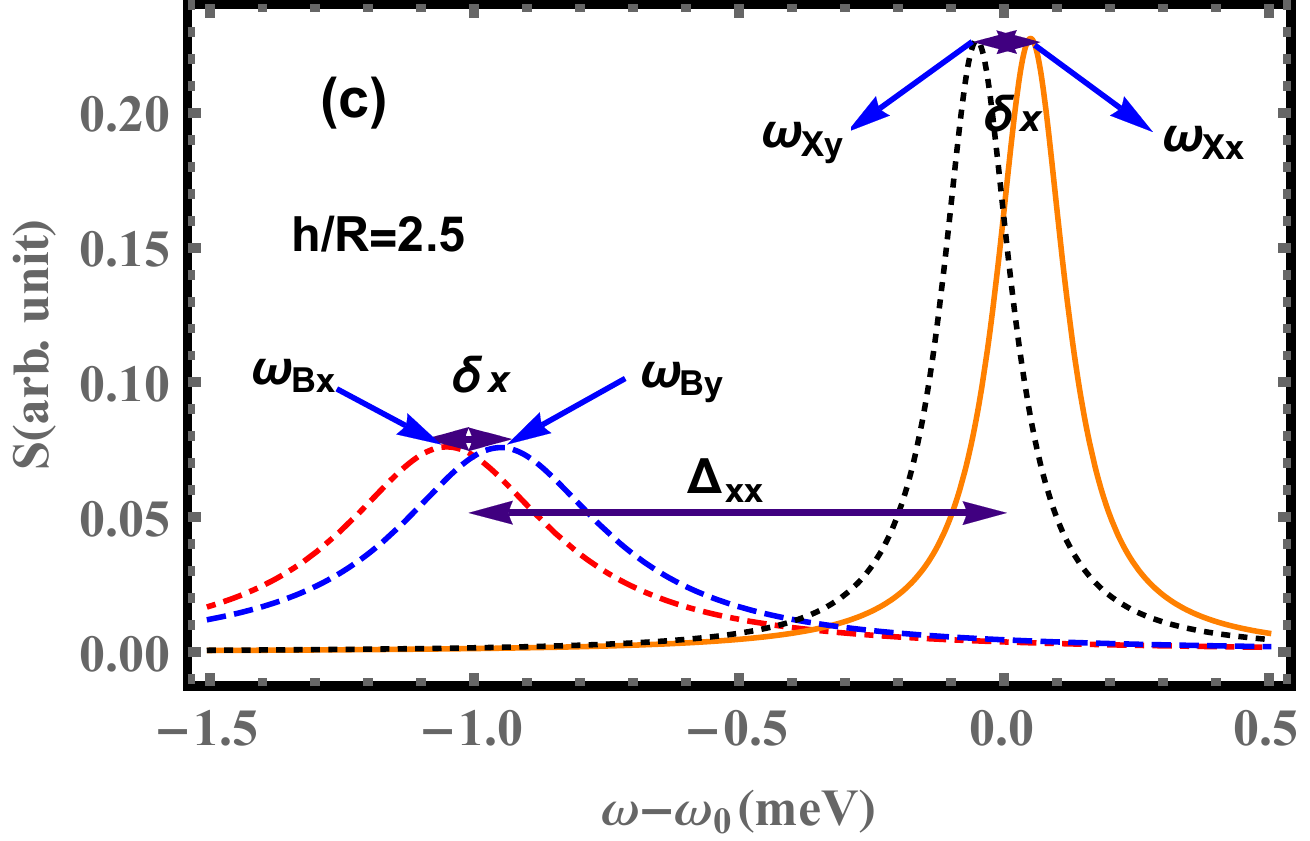}
        \includegraphics[width=\linewidth]{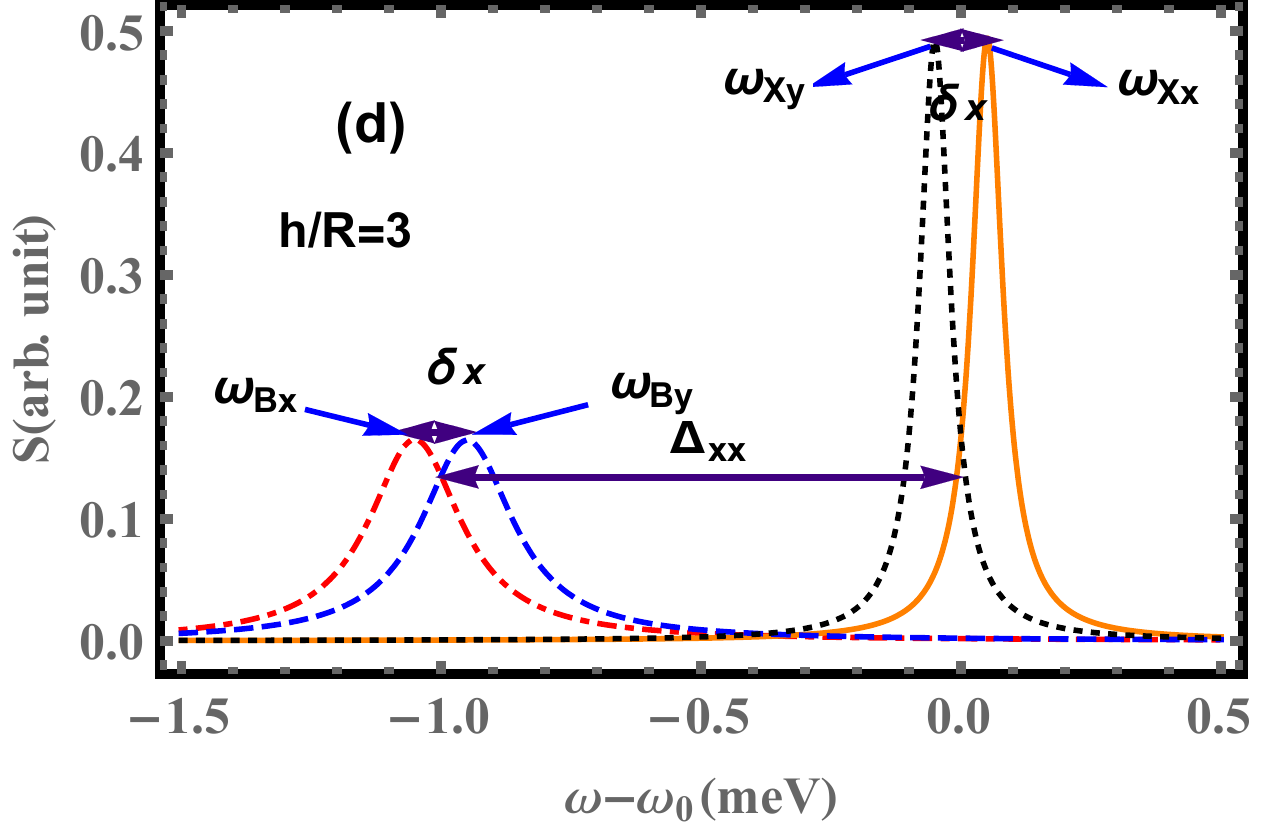}
        \caption{The spectrum of photon pairs generated through the cascade decay of a QD in the vicinity of a 7 nm radius MNP for $\Delta _{p}^{x}=(1-0.01 i)meV{{,}_{{}}}\Delta _{p}^{y}=-(2-0.01 i)meV,$ ${{\Delta }_{xx}}={{1}_{{}}}meV$, and $\delta_x={{0.1}_{{}}}meV$. As is seen, by increasing the separation distance the overlapping and FWHM of the spectra of the two orthogonal-polarization photons decrease.}
    \end{figure}
    \par To examine the influence of the MNP radius on the spectrum of the emitted photons from the biexciton cascade transition, in Fig. (5) we have plotted the spectrum against $\omega-{\omega}_{0}$ for $R$=14 nm. It has been shown that by increasing the radius of the MNP, the QD experiences a less structured reservoir ( see Fig. (2)).  As discussed earlier, by weakening the reservoir structure, the decay rate of the QD gets smaller. Consequently, the FWHM of the spectrum
    becomes comparable to the FSS energy, leading to a decrease in the overlapping of the spectrum.
    A comparison between Fig. (4) and Fig. (5)  reveals the effects of the MNP radius and the QD-MNP distance  on the spectra of the emitted photon pairs. As can be seen, by increasing these two parameters the overlap between the wave functions of the two orthogonal-polarized photons is reduced and consequently we expect the polarization entanglement between the photon pairs to decrease. Generally, by enhancing the effective geometrical parameters in the hybrid system, i.e., the QD-MNP separation distance and the radius of the MNP, the overlapping of the spectrum decreases (Figs. 4(b-d) and 5(d-d)). We emphasize that the FWHM broadening also has an impact on the concurrence. The broader the FWHM\texttt{\emph{,}} the better the concurrence.
    \begin{figure}
        \includegraphics[width=\linewidth]{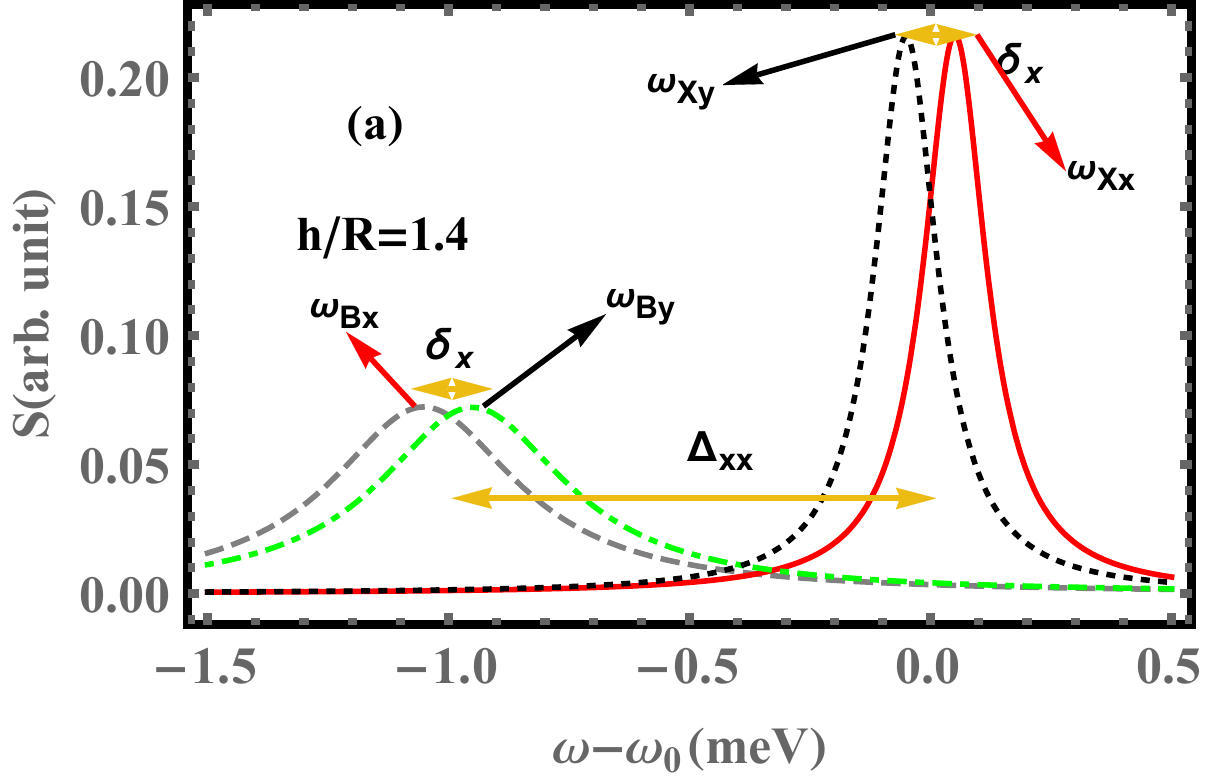}
        \includegraphics[width=\linewidth]{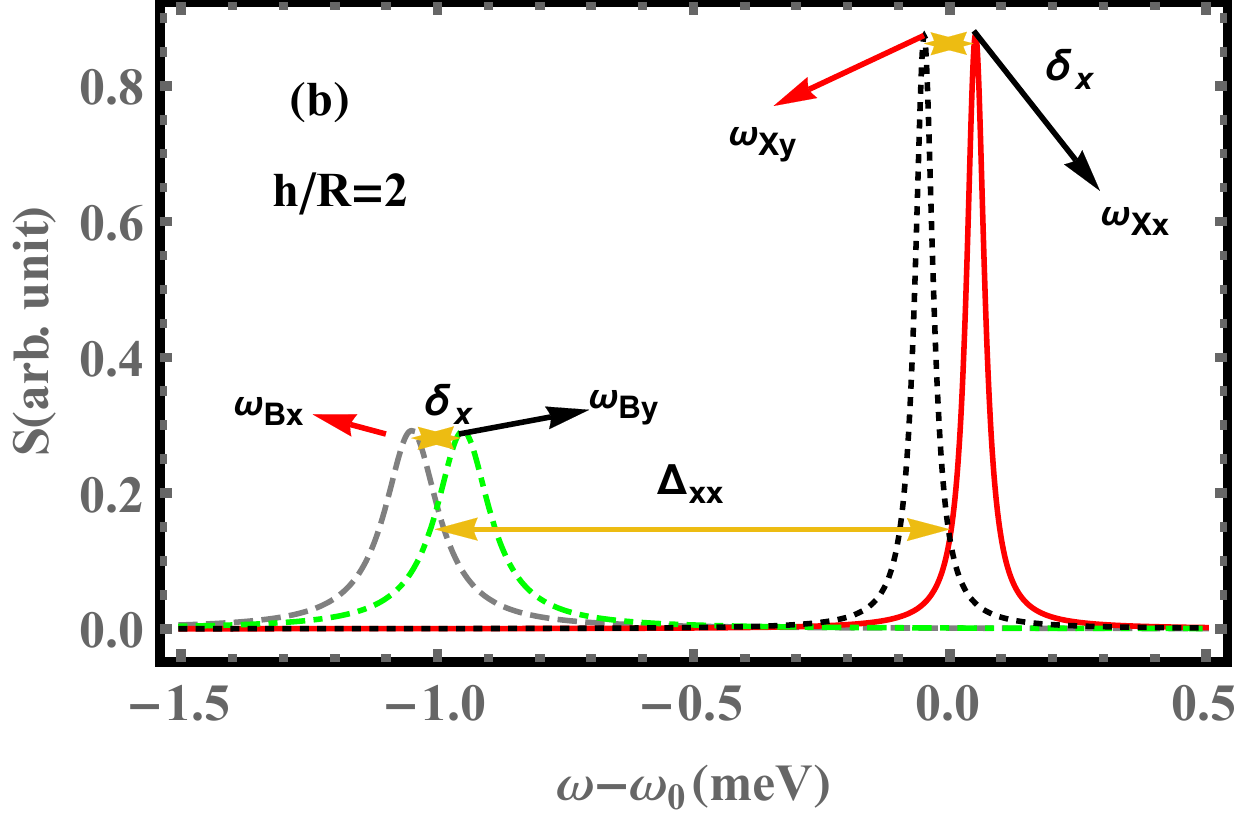}
        \includegraphics[width=\linewidth]{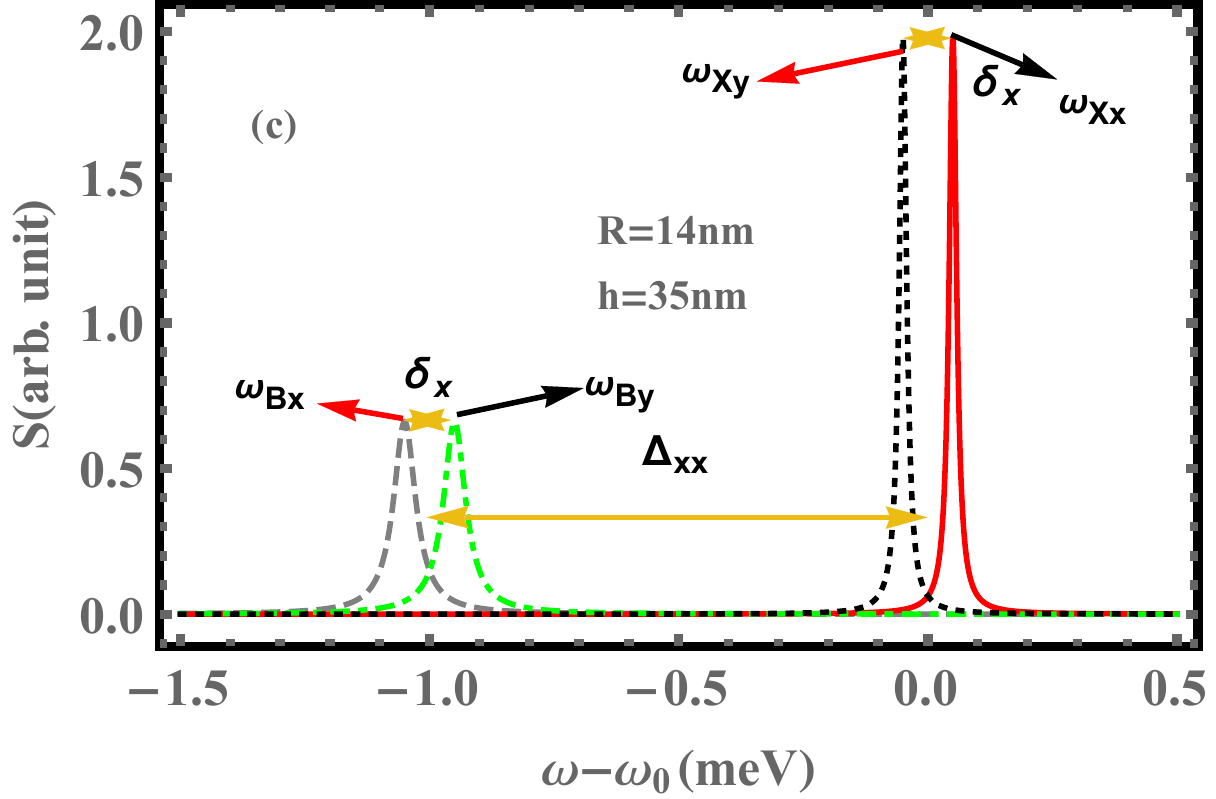}
        \includegraphics[width=\linewidth]{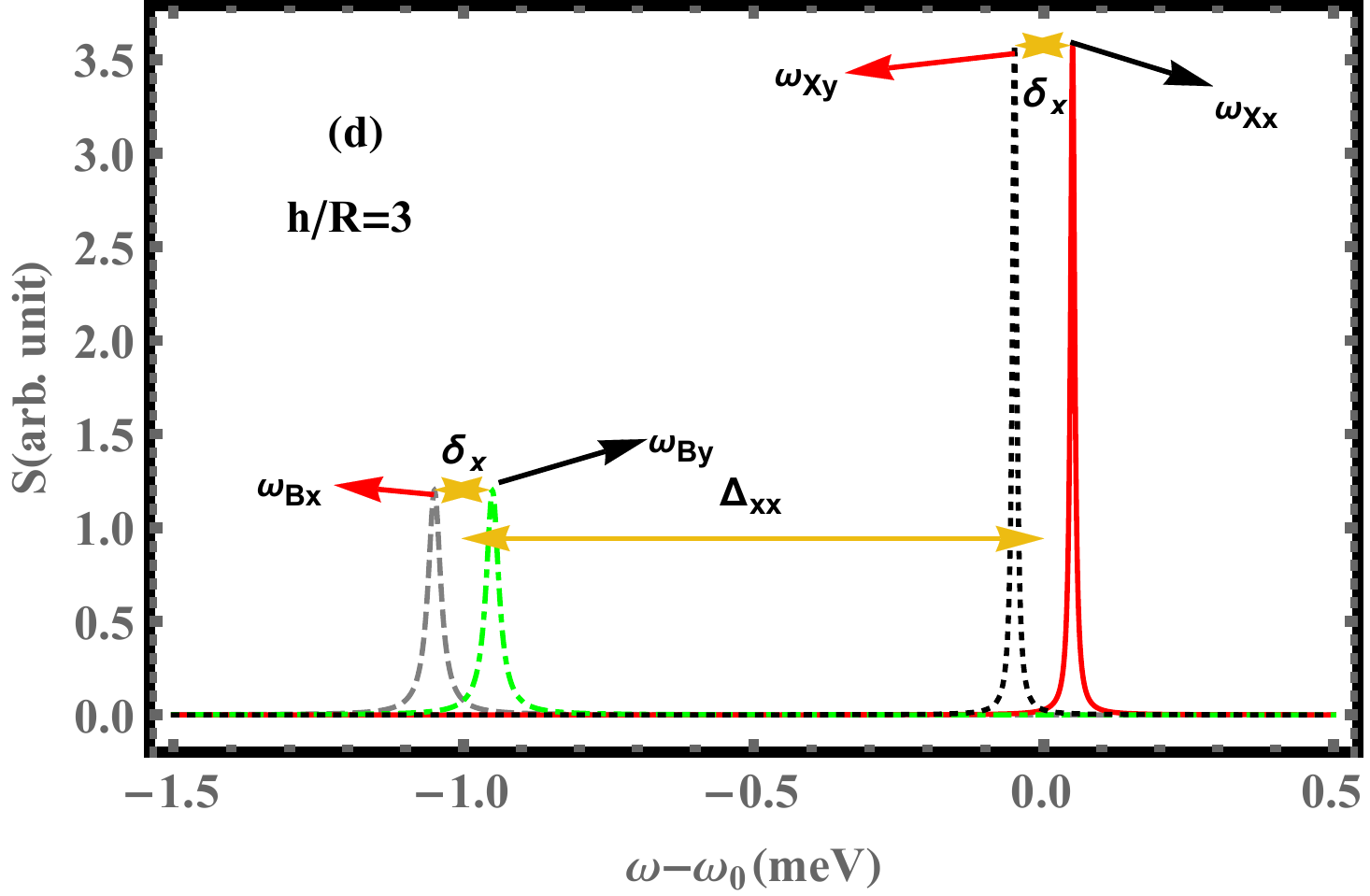}
        \caption{The spectrum of photon pairs generated through the cascade decay of a QD in the vicinity of a 14 nm radius MNP. The parameters are the same as those in Fig. (4). }
    \end{figure}
    \subsection{Polarization entanglement of the generated photon pairs}
    \par     The polarization entanglement can be distilled by using two frequency filters centered at the mean energies of photon pairs generated via the biexciton-exciton and the exciton-ground state transitions. Mathematically, the windows are determined by the projection operator and can be written as \cite{22}:
    \begin{align}\label{eq:17}
    W=\left\{ \begin{array}{lcr}
    1\texttt{\emph{,}} & \mbox{if $|\omega -{{\omega }_{0}}|<w$ }  \\
    1\texttt{\emph{,}} & \mbox{if $|\omega -{{\omega }_{0}}+{{\Delta }_{xx}}|<w $ }  \\
    0\texttt{\emph{,}} & otherwise  \end{array} \right.
    \end{align}
    Here, $w$ is the width of the spectral window centered at zero and $-\Delta_{xx}$, respectively, for photon pairs emitted in the biexciton and exciton radiative decays (see Fig.(4)).
    \begin{figure}
        \includegraphics[width=\linewidth]{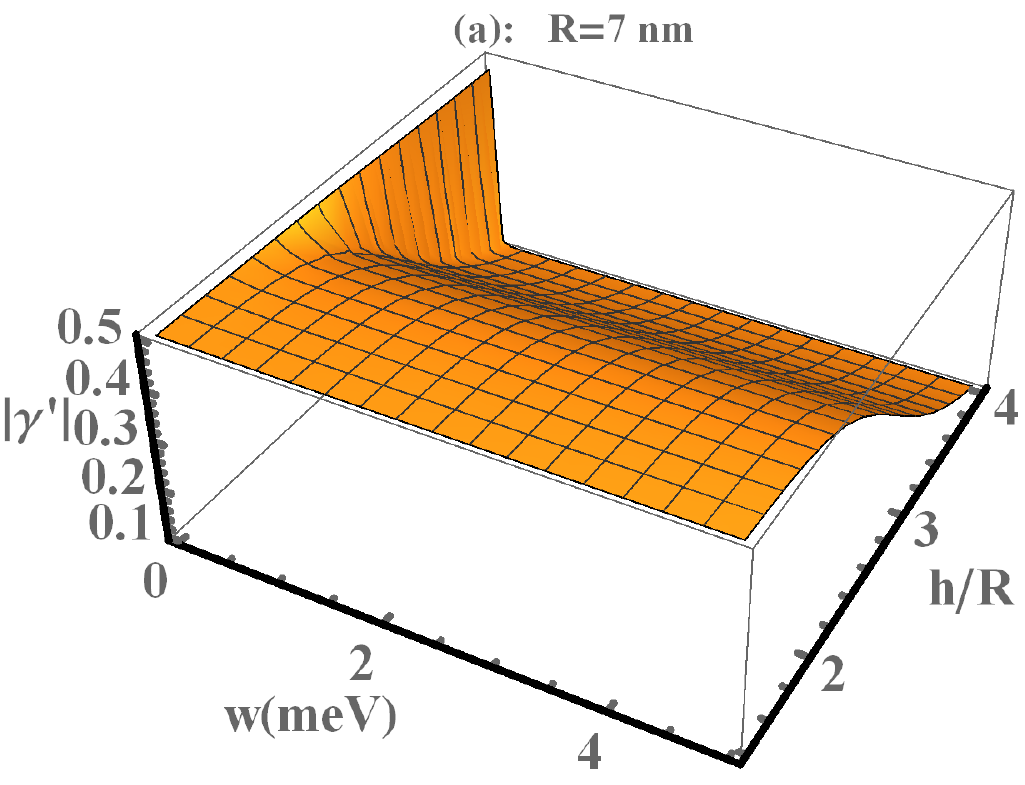}
        \includegraphics[width=\linewidth]{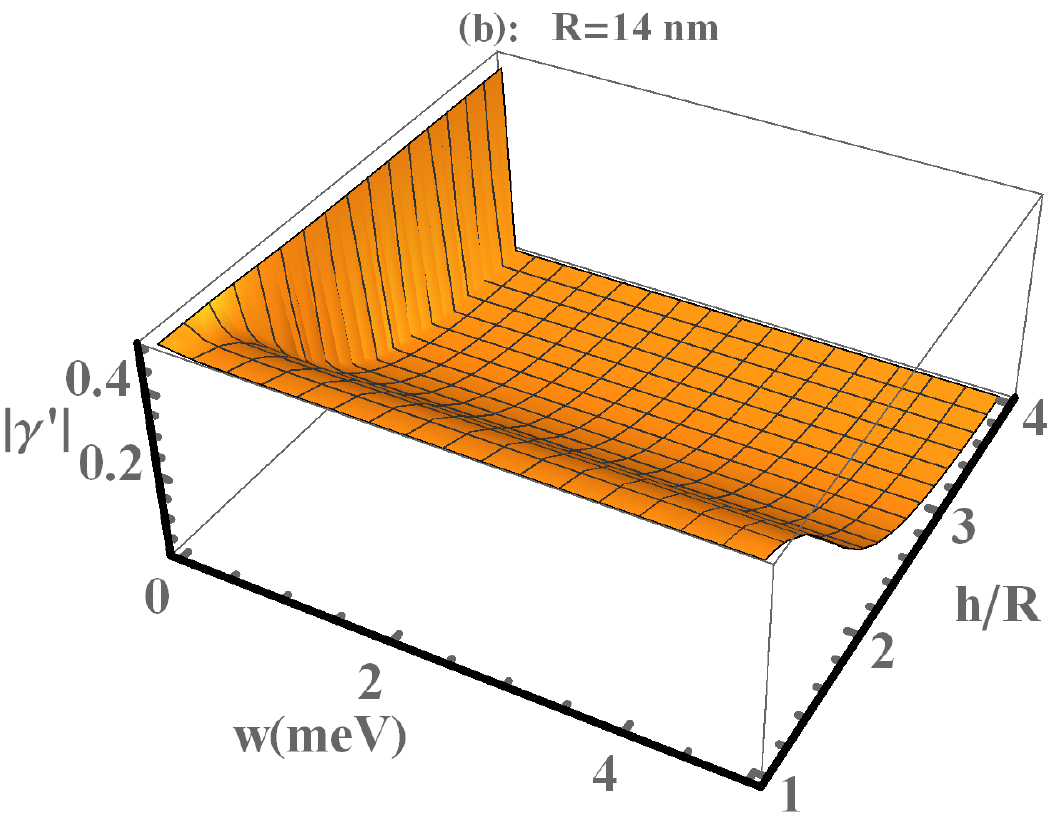}
        \caption{The polarization entanglement of the emitted photon pairs versus the filter width and ratio $h/R$ for a 7 nm radius MNP (a) and a 14 nm radius MNP (b). The other parameters are the same as those in Fig. (4).}
    \end{figure}
    \par
    To examine the influence of the action of filtering and the two important geometrical parameters, i.e.\texttt{\emph{,}} the QD-MNP separation distance and the radius of the MNP on the polarization entanglement for the state of the filtered photon pair, in Fig. (6) we have plotted the quantity $|{\gamma}'|$\texttt{\emph{,}} given by Eq. (\ref{eq:15})\texttt{\emph{,}} versus the filter width and the ratio of the QD-MNP separation distance to the radius of the MNP ($h/R$), for two different values of the MNP radius. As stated in section II, to ensure
    the Markovian behavior of the system the ratio of $h/R$ for a 7 nm radius MNP must be greater than 1.4 (Fig. (6a)), and for a 14 nm radius MNP it must be greater than 1 (Fig. (6b)). As can be seen, by increasing the ratio $h/R$ the polarization entanglement decreases in both cases. When the filter width decreases to an amount less than the FSS energy, the filtering operation has a significant role in the polarization entanglement for higher values of the ratio $h/R$. By comparing Fig.(6a) with Fig.(6b) we can easily recognize that with increasing the MNP radius the effect of the filtering on the polarization entanglement of photon pairs becomes more significant. Moreover\texttt{\emph{,}} for the smaller MNP the maximal preservation of polarization entanglement in higher ratio of $h/R$ is possible and at the equal ratio of $h/R$ the value of polarization entanglement for the smaller radius MNP is higher than the larger one. Finally, for the ratio $h/R=4$, the polarization entanglement value for the 14 nm radius MNP reaches to zero at a faster rate than for the 7 nm radius MNP.
    \par When the QD is located at a closer distance from the smaller MNP ($h/R=1.4$)\texttt{\emph{,}} the two polarized photons become indistinguishable by the almost complete overlapping of the x-polarized and y-polarized photons spectra generated in the biexciton-exciton transition. Indistinguishability of photons leads to the almost maximum value of the polarization entanglement ($|{\gamma}'|\cong1/2 $) showing that the filtering width has no effect on the polarization entanglement value. In other words, when the FWHM is much higher than the FSS energy, the filtering width, even smaller than the FSS, has no significant effect on the polarization entanglement value as shown in Fig. (4a). By increasing the ratio of $h/R$, the overlapping of the spectrum decreases (Figs. 4(b-d)) and thus the concurrence slakes. In this case, the value of polarization entanglement gets higher by making the frequency window narrower than the FSS energy because the narrower filter blocks the photons which are not overlapped.
    \begin{figure}
    	\includegraphics[width=\linewidth]{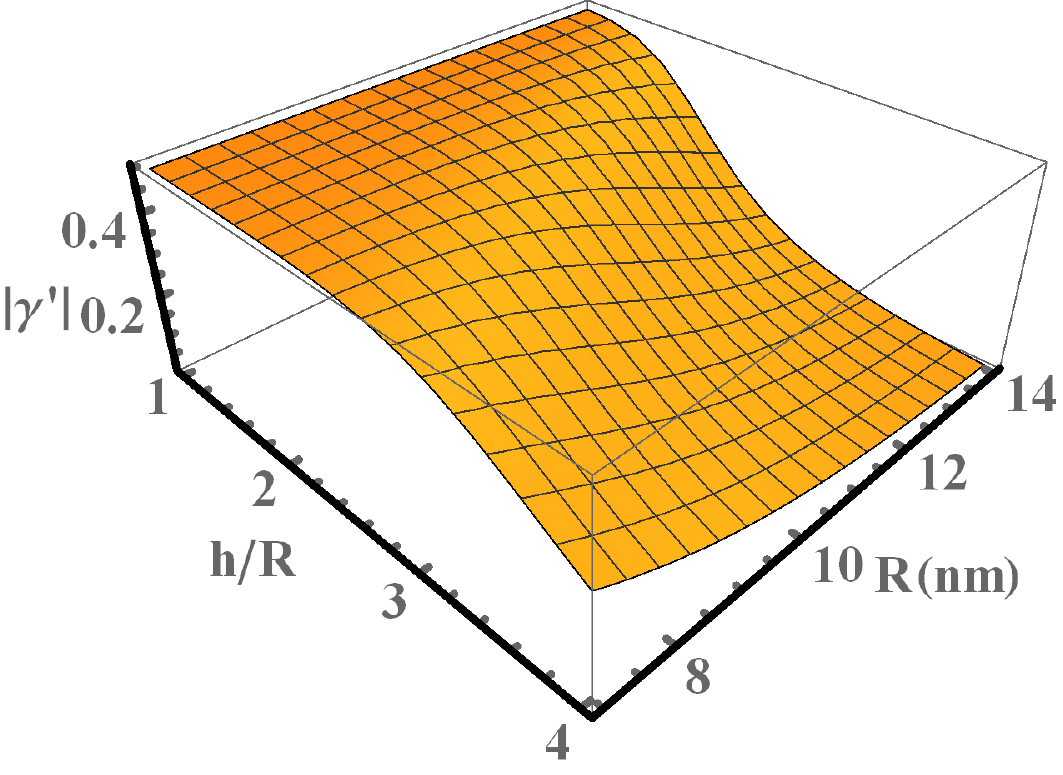}
    	\caption{The polarization entanglement of the emitted photon pairs versus the radius of the MNP and ratio $h/R$, for a fixed filter linewidth w=1meV.}
    \end{figure}
    \par 
    In order to investigate the existence of an optimal geometry including the radius of the MNP and the QD-MNP separation distance, we have quantified the entanglement by two criteria for the fixed filter line with $w=1 meV$: 
   \par 1)	We fixed the ratio of the QD-MNP separation distance to the radius of the MNP ($h/R$) and increased the radius of the MNP. We have plotted the quantity $|{\gamma }'|$ , given by Eq. (15) versus the radius of the MNP and the ratio of the QD-MNP separation distance to the radius of the MNP ($h/R$). As Fig. (7) shows, by increasing the ratio $h/R$, the polarization entanglement decreases for all values of the MNP radius. Given the same ratio of $h/R$, the smaller MNPs are more capable of preserving the polarization entanglement than the larger ones. Within this criterion, the smaller the MNP, the better the entanglement.
    \begin{figure}
    	\includegraphics[width=\linewidth]{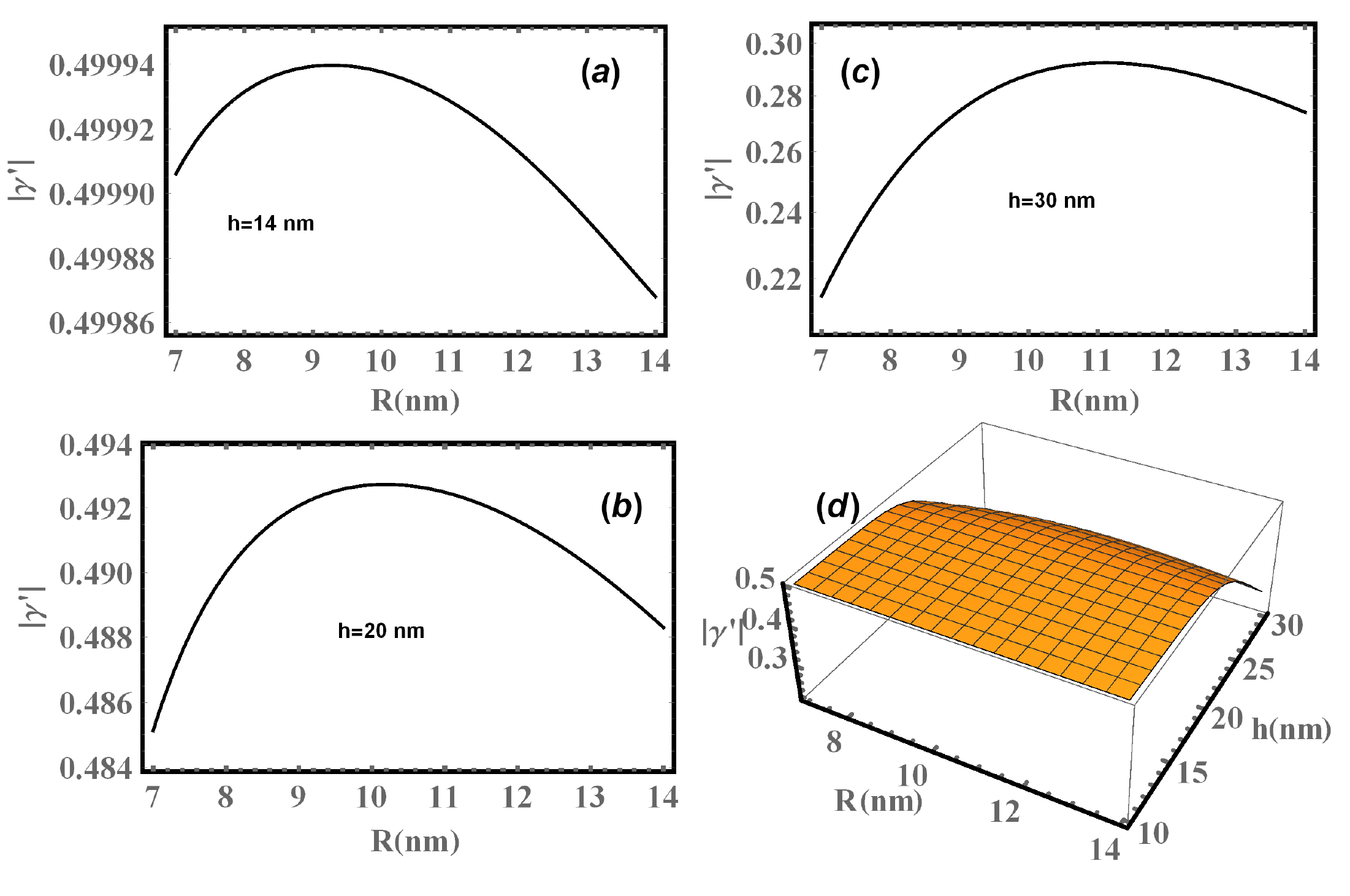}
    	\caption{The polarization entanglement of the emitted photon pairs versus the radius of the MNP for a fixed QD-MNP separation distance: (a): $h=14 nm$, (b): $h=20 nm$, (c): $h=30 nm$, and (d): The polarization entanglement of the emitted photon pairs versus the radius of the MNP and the QD-MNP separation distance $h$.}
    \end{figure} 
   \par
    2)	In Fig. (8), as we fix the QD-MNP separation distance (for example $h=14 nm$ in Fig.(8a)) and increase the MNP radius, the concurrence for a 9.5 nm MNP radius is optimum. The concurrences of the 7 nm, 9.5 nm, and 14 nm radiuses MNP are 0.4999, 0.49994, and 0.49987, respectively. The ratio of the QD-MNP separation distance to the radius of the MNP ($h/R$) is 2 for the first case, 1.47 for the second case, and 1 for the last one.
   When the QD-MNP separation distance is fixed to $h=20 nm$, the concurrence for a 10 nm MNP radius is optimum (Fig.(8b)). The concurrences of the 7 nm, 10 nm, and 14 nm radiuses MNP are 0.4845, 0.492, and 0.488, respectively. The ratio of the QD-MNP separation distance to the radius of the MNP ($h/R$) is 2.86 for the first case, 2 for the second case, and 1.43 for the last one.
   In Fig (8c) by increasing the QD-MNP separation distance to $h=30 nm$, the concurrence for a 11 nm MNP radius is optimum. In this case, the differences between maximum and minimum values of concurrences increase. The three-dimensional plot of the concurrence versus the radius of MNP and the QD-MNP separation distance is plotted in Fig. (8d).
   Within this criterion, the optimum radius of the MNP depends on the QD-MNP separation distance (h). In general, by altering h from 14 nm to 30 nm, the optimum radius is approximately changed from 9.5 nm to 11 nm.
   Our focus is to show that the filtering has no effect on the amount of the polarization entanglement of this configuration, unless the width of the filter is smaller than the FSS energy. In this study, we have shown that the polarization entanglement can be preserved in a long distance from the small MNP without filtering. Of course, if a filter with narrower width than the FSS is used, we can achieve a highly entangled pair of photons.

    \section{CONCLUSION}
    	 \par      We have proposed an approach to overcome the FSS for generating highly entangled photon pairs in the QD-MNP hybrid system within the Markovian dynamics. The degree of the polarization entanglement is quantified by the concurrence and is controlled
    	 by the geometrical parameters of the hybrid system\texttt{\emph{,}} that is\texttt{\emph{,}} the separation distance between the QD and MNP as well as the radius of MNP.
    	 According to our results\texttt{\emph{,}} the presence of a QD in the vicinity of a MNP will improve the entangled photon generation from the QD. In the system under consideration\texttt{\emph{,}} the maximum value of polarization entanglement is accessible at the expense of increasing the FWHM of the x-polarized and y-polarized spectra.
    	 \par In references \cite{22,24,27,28}\texttt{\emph{,}} a four-level QD embedded within an anisotropic photonic crystal is studied as the source of polarized entangled photons by making the intermediate exciton states of the QD degenerate in strong coupling regime. It is widely accepted that achieving a strong coupling regime is often more difficult than achieving a weak coupling regime. The contribution of the present study is the introduction of a system that can be realistically achieved through experiments using a hybrid system composed of a QD in the vicinity of a MNP even in a weak coupling regime. To the best of our knowledge\texttt{\emph{,}} it is the first time that such a study, concerning polarized entangled photon pair emission from a QD in the proximity of a MNP  is reported.
\bibliography{apssamp}

\end{document}